%% file: main.tex
\documentclass[12pt,letterpaper,utf8]{article}
\usepackage{amsmath,amsthm,cancel,eurosym,geometry,ulem,graphicx,caption,color,comment,caption,pdflscape,array,mathtools,multirow,hhline,bm,bbm,enumerate,optidef,nameref,subcaption,import,placeins,booktabs,microtype}
\usepackage{thmtools,thm-restate}
\usepackage{setspace}
\usepackage[multiple]{footmisc}
\usepackage[dvipsnames,table,xcdraw]{xcolor}

\usepackage{tikz}
\usetikzlibrary{shapes.geometric, arrows, fit, positioning}
\usetikzlibrary{calc}
\usetikzlibrary {shapes.geometric}

\numberwithin{equation}{section}

\usepackage{enumitem}
    \setlist[itemize]{noitemsep,nolistsep}
    \setlist[enumerate,1]{noitemsep,nolistsep,label=(\arabic*)}
    \setlist[enumerate,2]{noitemsep,nolistsep,label=(\alph*)}
\usepackage{titlesec}

\usepackage{titling}
\usepackage[page,title,toc]{appendix}

\usepackage[T1]{fontenc}
\usepackage[utf8]{inputenc}
\usepackage{amsfonts}
\usepackage{amssymb}

\usepackage{accents}
\newcommand{\ubar}[1]{\underaccent{\bar}{#1}} 

\usepackage[nottoc,numbib]{tocbibind}
\usepackage[backend=biber,style=authoryear-comp,maxbibnames=30,citetracker=true,maxcitenames=2,uniquename=false,uniquelist=false,natbib]{biblatex}
\AtEveryBibitem{
  \clearfield{issn}
  \clearfield{isbn}
}
\bibliography{citations_uniform}
\newrobustcmd*{\citefirstlastauthor}{\AtNextCite{\DeclareNameAlias{labelname}{given-family}}\citet*}

\makeatletter
\newcommand{\upmax}{\def\blx@maxcitenames{99}}
\newcommand{\dnmax}{\def\blx@maxcitenames{1}}
\makeatother

\theoremstyle{plain}

\theoremstyle{definition}

\theoremstyle{remark}

\AtBeginEnvironment{proof}{\setcounter{case}{0}}
\AtEndEnvironment{case}{$\blacktriangle$}
\AtBeginEnvironment{proof}{\setcounter{subcase}{0}}
\AtEndEnvironment{subcase}{$\blacktriangle$}
\AtBeginEnvironment{proof}{\setcounter{step}{0}}

\AtBeginEnvironment{appendices}{\crefalias{section}{appendix}}
\AtBeginEnvironment{appendices}{\crefalias{subsection}{appendix}}
\AtBeginEnvironment{appendices}{\crefalias{subsubsection}{appendix}}

\newcolumntype{L}[1]{>{\raggedright\let\newline\\arraybackslash\hspace{0pt}}m{#1}}
\newcolumntype{C}[1]{>{\centering\let\newline\\arraybackslash\hspace{0pt}}m{#1}}
\newcolumntype{R}[1]{>{\raggedleft\let\newline\\arraybackslash\hspace{0pt}}m{#1}}

\let\phi\varphi
\usepackage{etoolbox}

\usepackage[hyperfootnotes=false]{hyperref}
\hypersetup{
	linktoc=page,
	unicode=false,          
	pdftoolbar=true,        
	pdfmenubar=true,        
	pdffitwindow=false,     
	pdfstartview={Fit},    
	pdftitle={},    
	pdfauthor={},     
	pdfcreator={},   
	pdfproducer={}, 
	pdfkeywords={}, 
	pdfnewwindow=true,      
	colorlinks=true,       
	linkcolor=MidnightBlue,          
	citecolor=MidnightBlue,        
	filecolor=black,      
	urlcolor=BrickRed,		
	anchorcolor=black,
	menucolor=black,
	runcolor=black,
}

\usepackage[nameinlink]{cleveref}
	\crefname{theorem}{theorem}{theorems}
	\Crefname{theorem}{Theorem}{Theorems}
	\Crefname{lemma}{Lemma}{Lemmata}
	\crefname{lemma}{lemma}{lemmata}
	\crefname{proposition}{proposition}{propositions}
	\Crefname{proposition}{Proposition}{Propositions}
	\crefname{definition}{definition}{definitions}
	\Crefname{definition}{Definition}{Definitions}
	\crefname{corollary}{corollary}{corollaries}
	\Crefname{corollary}{Corollary}{Corollaries}
	\crefname{assumption}{assumption}{assumptions}
	\Crefname{assumption}{Assumption}{Assumptions}
	\Crefname{myobservation}{Observation}{Observations}
	\Crefname{claim}{Claim}{Claims}
	\Crefname{table}{Table}{Tables}
	\Crefname{tabular}{Table}{Tables}
	\crefname{appsec}{appendix}{appendices}
	\Crefname{appsec}{Appendix}{Appendices}
	\Crefname{case}{Case}{Cases}
	\Crefname{subcase}{Subcase}{Subcases}
	\Crefname{example}{Example}{Examples}
	\Crefname{remark}{Remark}{Remarks}
  \Crefname{footnote}{Footnote}{Footnotes}
  \crefname{footnote}{footnote}{footnotes}
  \Crefname{appendices}{Appendix}{Appendices}
  \crefname{appendices}{appendix}{appendices}
  \Crefname{figure}{Figure}{Figures}

\usetikzlibrary{patterns}
\usepackage{xcolor}

\definecolor{darkgreen}{rgb}{0.0, 0.5, 0.0}
\definecolor{darkred}{rgb}{0.5, 0.0, 0.0}

\usepackage{tcolorbox}

\tcbuselibrary{breakable}
\tcbset{
  width=1\textwidth,
  halign=justify,
  center,
  breakable,
  colback=white    
}

\newtcolorbox[auto counter, crefname={Model Box}{Model Boxes}]{modelbox}[2][]{%
colback=white, halign=justify, center, width=1\textwidth, breakable, fonttitle=\bfseries,
title=Model Box~\thetcbcounter: #2,#1}

\makeatletter
\newbibmacro*{related:bytranslator}[1]{%
  \entrydata{#1}{%
    \renewbibmacro*{name:hook}[1]{%
      \ifnumequal{\value{listcount}}{1}
        {\begingroup
         \mkrelatedstring%
         \lbx@initnamehook{#1}%
         \endgroup}
        {}}%
    \printnames[bytranslator]{translator}%
    \clearname{translator}%
    \setunit*{\addspace\bibstring[\mkrelatedstring]{astitle}\space}%
    \usebibmacro{title}%
    \ifentrytype{article}
      {\usebibmacro{in:}%
       \usebibmacro{journal+issuetitle}}
      {}
    \ifentrytype{inbook}
      {\usebibmacro{in:}%
       \usebibmacro{bybookauthor}%
       \newunit\newblock
       \usebibmacro{maintitle+booktitle}%
       \newunit\newblock
       \usebibmacro{byeditor+others}%
       \newunit\newblock
       \printfield{edition}}
      {}
    \ifentrytype{incollection}
      {\usebibmacro{in:}%
       \usebibmacro{maintitle+booktitle}%
       \newunit\newblock
       \usebibmacro{byeditor+others}%
       \newunit\newblock
       \printfield{edition}%
       \newunit
       \iffieldundef{maintitle}
         {\printfield{volume}%
          \printfield{part}}
         {}}
      {}
    \ifentrytype{inproceedings}
      {\usebibmacro{in:}%
       \usebibmacro{maintitle+booktitle}%
       \newunit\newblock
       \usebibmacro{event+venue+date}%
       \newunit\newblock
       \usebibmacro{byeditor+others}%
       \newunit\newblock
       \iffieldundef{maintitle}
         {\printfield{volume}%
          \printfield{part}}
         {}}
      {}
    \setunit{\addspace}%
    \printtext[parens]{%
      \printlist{location}%
      \iflistundef{publisher}
        {\setunit*{\addcomma\space}}
        {\setunit*{\addcolon\space}}%
      \printlist{publisher}%
      \setunit*{\addcomma\space}%
      \printdate}
      \newunit\newblock
      \printfield{note}%
      \newunit\newblock
      \printfield{chapter}%
      \setunit{\bibpagespunct}%
      \printfield{pages}}}
\makeatother

\defcitealias{galileo1609doge}{Galileo (1609)}
\defcitealias{galileo1610vinta}{Galileo (1610)}

\title{Designing Scientific Grants\thanks{We thank Ben Jones and participants at the NBER EIPE Conference, 2024 for stimulating comments. Marco Ottaviani gratefully acknowledges financial support from MUR-PRIN 2022 (Prot. 2022RL4WMT, ``Finanziamento dell’Unione Europea – NextGenerationEU''). Justus Preusser gratefully acknowledges financial support from the European Research Council (HEUROPE 2022 ADG, GA No. 101055295 – InfoEcoScience).}}

\author{
Christoph Carnehl\thanks{\protect Department of Economics and IGIER, Bocconi University, \textit{\href{mailto:christoph.carnehl@unibocconi.it}{christoph.carnehl@unibocconi.it}}.}
\and
Marco Ottaviani\thanks{\protect Department of Economics and IGIER, Bocconi University, \textit{\href{mailto:marco.ottaviani@unibocconi.it}{marco.ottaviani@unibocconi.it}}.}
\and
Justus Preusser\thanks{\protect Department of Economics and IGIER, Bocconi University, \textit{\href{mailto:justus.preusser@unibocconi.it}{justus.preusser@unibocconi.it}}.}
}

\date{\today}

\begin{document}

\begin{titlepage}
\maketitle
\vspace{-0.5cm}
\begin{abstract}
    \input{./abstract.tex}
\end{abstract}
\end{titlepage}

\input{./introduction}
\input{./why_fund_science_with_grants}

\input{./goal_and_challenges}
\input{./grant_design}

\input{./concluding_remarks}
\printbibliography

\end{document}

%% file: abstract.tex
This paper overviews the economics of scientific grants, focusing on the interplay between the inherent uncertainty in research, researchers' incentives, and grant design. Grants differ from traditional market systems and other science and innovation policy tools, such as prizes and patents. We outline the main economic forces specific to science, noting the limited attention given to grant funding in the economics literature. Using tools from information economics, we identify key incentive problems at various stages of the grant funding process and offer guidance for effective grant design.

In the allocation stage, funders aim to select the highest-merit applications while minimizing evaluation costs.
The selection rule, in turn, impacts researchers' incentives to apply and invest in their proposals. 
In the grant management stage, funders monitor researchers to ensure efficient use of funds. We discuss the advantages and potential pitfalls of (partial) lotteries and emphasize the effectiveness of staged grant design in promoting a productive use of grants.

Beyond these broadly applicable insights, our overview highlights the need for further research on grantmaking. Understudied areas include, at the micro level, the interplay of different grant funding stages, and at the macro level, the interaction of grants with other instruments in the market for science.

%% file: introduction.tex
Scientific grants---upfront payments to support research in promising yet uncertain areas---are a key source of research funding.\footnote{For example, the National Institutes of Health (NIH) in the US had a budget of around $45$ billion dollars in 2022 (\url{https://www.nih.gov/about-nih/what-we-do/budget}). The Horizon Europe program has a budget of around $95$ billion euros for the period 2021-2027 (\url{https://commission.europa.eu/funding-tenders/find-funding/eu-funding-programmes/horizon-europe_en}).} In this paper, we explore the economics of grants, with an emphasis on the interplay between the uncertainty inherent in research, incentives and interests of both researchers and funders, and the design of grant schemes. The following historical account from the earliest days of modern science illustrates many of the key issues that prevail until today.

In the summer of 1609, a relatively obscure lecturer of Mathematics at the University of Padua by the name of Galileo Galilei \parencite*{galileo1609doge} wrote a letter to the Doge of Venice to present 
``a new invention of a telescope
[...] to bring extraordinary benefit to Your Highness.'' In actuality, Galileo did not invent but only enhanced the telescope. 
Nevertheless, he hinted that he would relinquish to the Doge the power to control the further diffusion of telescopes,
a dubious claim given the quick diffusion of the tool as a fashionable toy among well-off Venetians. As chronicled by \citet{biagioli2019galileo}, Galileo obtained a salary increase and tenure at the University of Padua. 

In the following months, Galileo pointed the telescope at the sky, making a series of astronomical discoveries---among these, the four largest satellites of Jupiter---that marked the dawn of modern science.
Galileo hastened to establish priority and published his treatise \textit{Sidereus Nuncius} in March 1610, only two months after his discovery.\footnote{Urgency was justified given the rapid diffusion of the telescope. As a matter of fact, Galileo was later involved in a bitter priority dispute with German astronomer Simon Marius, who apparently observed the four satellites of Jupiter one day after Galileo \citep{pasachoff2015simon}.} 

Galileo seized the opportunity for a major career advancement and a return to his native Tuscany. He dedicated his treatise to
Cosimo de’ Medici, the Grand Duke of Tuscany, and named the four satellites of Jupiter that he discovered the Medicean Stars. 
Galileo \parencite*{galileo1610vinta} negotiated a prestigious appointment, desiring that ``in addition to the title of mathematician His Highness will annex that of philosopher,'' 
and sought the financial support that would enable him to focus on research ``[...] because giving private lessons and taking scholars as boarders constitute something of an obstacle[.]''
Galileo highlighted the prestige his patron would gain,
promising ``[...] such inventions as no other prince can match, for of these I have a great many and am certain I can find more.''

Galileo also recognized his value as a peer evaluator: ``Concerning these inventions which belong to my calling, His Highness may rest assured that he will not be wasting his money on them, as perhaps he has done at other times in great quantity, nor will he miss out on any that are useful and good which have been proposed to him by other men.'' 

Moved by this eloquent appeal, Cosimo appointed Galileo chief mathematician and philosopher to the Grand Duchy of Tuscany in the summer of 1610.
Simultaneously, Galileo obtained the position of head mathematician of the University of Pisa, carrying no obligation to reside or teach there---essentially a full teaching buyout.

Although four hundred years have passed since the time of Galileo, his entre\-preneurial efforts to secure funding and recognition seem to perfectly anticipate the labors of modern researchers. Like Galileo, researchers appeal to funders by demonstrating a record of successes while promising impactful new discoveries.
Like the Doge and the Grand Duke, funders want to channel resources towards talented researchers and rely on researchers' expertise to evaluate funding opportunities. And like Galileo, researchers are under pressure to publish quickly and establish priority.

Unlike traditional market systems widely analyzed by economists, scientific grants are characterized by upfront payments without contractible goals, giving researchers significant autonomy in using these resources. This flexibility is essential due to the nature of grant-supported research, which is often open-ended and uncertain. For example, it is typically impossible to define the value or outcome of research in advance, given its inherently open-ended and exploratory nature. Neither funding agencies nor researchers can anticipate whether years of research will ever lead to a significant discovery, whether the scientific community will realize the significance, or whether the discovery will have an impact beyond the scientific community. Driven by passion, intellectual curiosity, and the hope for scientific recognition, investigators conduct “blue-sky” research to solve open puzzles and challenge conventional wisdom. More than perhaps any other source of funding, grants thus leverage researchers' self-motivation for discovery and recognition within the community.

Compared to other funding instruments, such as patents and research prizes, grants seem to have received considerably less attention in the economics literature. 
We are confronted with at least two questions. 

\begin{figure}[ht]
    \centering
    \begin{tikzpicture}[node distance=1.75cm,scale=1]
    \node (grants) [draw, rectangle] {Grants};
    \node (effort2) [shade, top color=lightgray!40, bottom color=lightgray!40, draw, below of=grants] {Research};
    \node (results) [anchor=top, draw, rectangle, below of=effort2, xshift=0cm] {Satisfaction, recognition, patents and prizes};
   
    \draw [->, line width=1pt] (effort2) -- (results);
    \draw [->, line width=1pt] (grants) -- (effort2);
    \end{tikzpicture}
    \caption{A schematic view of science funding.}
    \label{fig:intro_science_funding}
\end{figure}
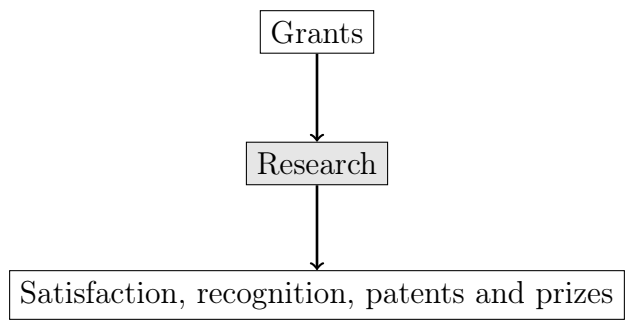

First, what economic reasons are there for funding science using grants instead of patents or research prizes?
\Cref{fig:intro_science_funding} offers a schematic view of science funding. 
Grants are awarded before researchers begin working on their projects.
Upon successfully completing research projects, researchers stand to gain intellectual satisfaction, recognition within the community, and patents and prizes.
Existing work has illuminated the impact of patents and prizes on researchers' incentives.
How do these incentives interact with the incentives provided by scientific grants? What are the relative strengths and weaknesses of these instruments?

\begin{figure}[ht]
\centering
\begin{tikzpicture}[node distance=1.75cm,scale=1]
\node (application) [shade, top color=lightgray!40, bottom color=lightgray!40, draw] {Application, investment and persuasion effort};

\node (prospective) [draw, rectangle, below of=application, yshift=0cm] {Prospective evaluation};
\draw [->, line width=1pt] (application) -- (prospective);

\node (allocation) [draw, rectangle, below of=prospective, yshift=0cm] {Funding};
\node (retrospective_evaluation) [draw, rectangle, left=1.75cm of allocation, text width=2.5cm, align=center] {Retrospective evaluation of past grants};

\draw [->, line width=1pt] (prospective) -- (allocation);
\draw [->, line width=1pt] (retrospective_evaluation) -- (allocation);

\node (effort) [shade, top color=lightgray!40, bottom color=lightgray!40, draw, below of=allocation] {Research};
\node (management) [draw, rectangle, below of=effort, yshift=0.6cm] {Grant management};
\node[fit=(effort) (management), ellipse, draw, dotted, minimum width=5cm, minimum height=2.75cm, inner sep=0pt] (ellipse) {};
\draw [->, line width=1pt] (allocation) -- (ellipse);
\draw [<->, line width=1pt] (effort) -- (management);

\node (results) [draw, rectangle, below of=management, xshift=0cm] {Satisfaction, recognition, promotion, patents and prizes};
\draw [->, line width=1pt] (ellipse) -- (results);

\end{tikzpicture}
\caption{A schematic view of grant funding.}
\label{fig:intro_grant_funding}
\end{figure}
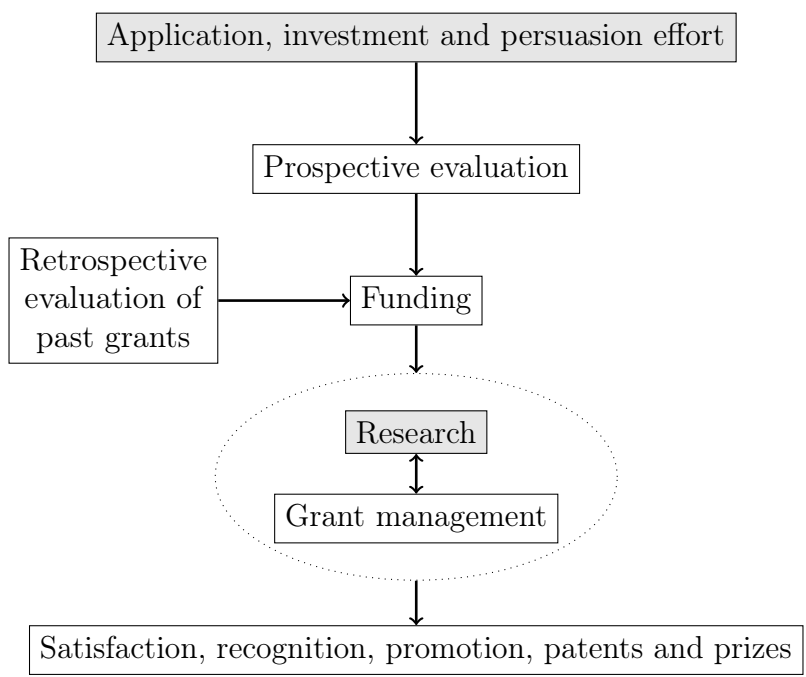

Second, if one uses grants, what should they look like, and how should they be allocated?
\Cref{fig:intro_grant_funding} gives a schematic view of the grant funding process.
The funding process involves a number of stages where either researchers (in the shaded boxes) or grantmakers (in the unshaded boxes) make decisions.
How does prospective evaluation impact the incentives of researchers to invest in the intrinsic quality or persuasiveness of their proposals?
What measures can grantmakers take to stimulate or mitigate these investments?
How can retrospective evaluation be used to effectively screen applicants and manage grantees?

We draw from information economics and mechanism design to gain insight into these issues.
Spread across a diverse set of papers, the literature has developed several important results that shed light on individual pieces from \Cref{fig:intro_grant_funding,fig:intro_science_funding}.
However, in our view, more remains to be done to develop a unifying theory on the relationship between different pieces, clarifying the interaction of different stages of grant funding (\Cref{fig:intro_grant_funding}) and the interaction between different stages of science funding (\Cref{fig:intro_science_funding}).

Our work complements recent overviews by \citet{price2019grants} and \citet{azoulay2020scientificgrantfunding}. \citet{price2019grants} explains the details of grant allocation at the National Institutes of Health (NIH) and offers a rebuttal against common critiques of grants. \citet{azoulay2020scientificgrantfunding} review the empirical economics literature on grants and discuss various aspects of practical grant design. 

The paper proceeds as follows.
In the next section, we briefly discuss the economic justification for science funding, and we delineate some of the situations in which grants appear more suitable than patents and research prizes.
The remainder of the paper focuses on grants.
In \Cref{sec:goals_and_challenges}, we lay out the basic incentive problems that arise in grant allocation.
In \Cref{sec:application_process,sec:allocation_rule,sec:postaward_management}, we then discuss in more detail the economics of the application and allocation processes, as well as the post-award management stage.
Finally, in \Cref{sec:supply_and_direction}, we ask how the supply of grants impacts the direction of research.

%% file: why_fund_science_with_grants.tex
\section{Why and how to fund science?}\label{sec:why_fund_using_grants}

Scientific research is central to the economic well-being of modern societies since it creates valuable new knowledge. As explained by \citet{Arrow1962}, however, there are several reasons why one may think that markets fail to incentivize an efficient production of knowledge. 
First, while scientific research consumes large amounts of resources, its output, knowledge, is non-rival since it can be reproduced at (almost) no cost. 
Second, knowledge production creates positive externalities by inspiring follow-up research.
Third, there is substantial uncertainty about the nature and value of research output.
It is difficult to insure against this uncertainty since one needs specialist knowledge to evaluate both a project's ex-ante prospects and, at the ex-post stage, the work that was actually carried out.

For these reasons, there is an economic justification for government interventions that adjust incentives in the market for knowledge production. Governments around the world recognize the importance of science policy for the prosperity of societies in general, but also as a strategic investment in national competitiveness and economic security \citep{dasgupta1994toward}. \citet{jones2021science} summarizes a large body of evidence that points to the large social gain from subsidizing science and innovation.

\subsection{Incentives in the production of knowledge}

Even accepting the need for policies to encourage knowledge production, it is not at all clear what these policies should be.
In practice, researchers' incentives are shaped by various monetary and non-monetary instruments, as illustrated in \Cref{fig:intro_science_funding}.

Grants act at the ex-ante stage and provide \emph{push incentives}.
Specifically, grants provide researchers with resources to carry out projects (in the form of specialized equipment and access to data, for example) and reduce the costs of conducting research (by funding personnel, for example).
Thus, grants \emph{push} researchers into a broad direction while leaving them with substantial leeway to choose the exact topic and method.
Since grants are awarded before any knowledge is produced, the allocation of grants relies on the costly ex-ante evaluation of proposals and researchers.

Patents and prizes offer rewards only for completed projects.
Thus, patents and prizes \emph{pull} researchers towards questions that the market or the prize-maker values. 
Accordingly, costly evaluation happens ex-post when inventors apply for patents \citep{matcham2023screening} or the prize committee chooses a winner.

In addition to these monetary incentives, there are non-monetary incentives at work, stemming from intellectual satisfaction and recognition in the scientific community from establishing priority.
\citet{merton1957priority} emphasizes the importance of the priority system for researchers' incentives.
\citet{hill2023scooped} find that narrowly failing to establish priority has a negative impact on publication and career outcomes for structural biologists.

\subsection{Why fund science using grants?}

We next delineate situations in which grants seem particularly useful.
There appear to be no formal economic analyses comparing grants to other funding instruments. Earlier work by \citet{wright1983economics,gallini2002intellectual} views grants as special fixed-price contracts and offers only a stylized analysis.

To begin with, we note that the non-monetary incentives stemming from priority and citations may be insufficient for encouraging an efficient production of knowledge.
The priority system and citations reward researchers proportionally to the impact of their work, which would seem to encourage work on important problems.
Alas, the system is imperfect.
The pressure to publish can have a detrimental impact.
When several researchers work on the same problem, there is a race to be the first to publish.
\citet{hill2024race} provide evidence that structural biology projects with higher recognition potential are completed faster and are of lower quality. 
Further, there is evidence of publication bias against null results (see \cite{andrews2019identification}, for example). The importance of citation measures in quantifying scientific impact and in promotion decisions can induce researchers to, possibly inefficiently, prioritize research fields that generate higher citation counts over others (see \cite{Olszewski2020citations}, for example).
Thus, even though non-monetary incentives surely play an important role, it is worth examining the role of monetary incentives.

Compared to patents and research prizes, grants have several strengths.
First and most straightforwardly, grants address financial constraints and allow funders to target specific people and institutions. Scientific apprenticeships, laboratory equipment, and scientific infrastructure are typically financed via grants \citep{azoulay2020scientificgrantfunding,price2019grants}. 
Thus, grants can direct resources towards minorities or compensate endowment differences across universities to level the playing field.\footnote{\citet{yang2022genderdiverse} show that more diversity in teams can lead to better research outcomes.}
Patents and prizes, offering rewards only for completed projects, arguably do less than grants to alleviate financial constraints.
While prizes can be targeted towards specific people and institutions, patents are typically untargeted.

Second, grants may be useful for aligning market values with social values.
A mismatch between the two values may arise when market participants are unwilling or unable to pay for research output; \citet{price2019grants,azoulay2020scientificgrantfunding} give the example of medical treatments whose social value far exceeds any individual's willingness to pay.
\citet{castillo2021market} indicate an enormous gap between the social and commercial incentives to expand vaccine capacity during the COVID-19 pandemic.
A mismatch between social and market values diminishes the appeal of funding instruments that, like patents, create incentives by linking the rewards from research to market returns.
\citet{bryan2017direction} provide theoretical support for the idea that patents fall short of guiding researchers to the right topics. 
In their framework, researchers choose both the intensity and the direction of their research. 
They show that, while patents can induce the efficient intensity of research, patents do not guarantee directional efficiency. One could conjecture that, here, grantmakers can step in by committing to identifying and supporting socially valuable topics.

Third, grants help address the uncertainty inherent in research \citep[see, for example,][for a discussion of uncertainty in research]{franzoni2023uncertainty}. 
Fundamental research often involves high uncertainty about the value of discoveries. Further, the commercial potential of discoveries may reveal themselves only much later.\footnote{See \citet{azoulay2019public,li2017applied,bryan2021innovation} for some evidence that grant-funded projects indeed lay the foundation for patented inventions.}
From theoretical work by \citet{aghion2008academic}, we can glean some intuition for the impact of this uncertainty on funding design. 
In their framework, researchers are willing to forgo high salaries in exchange for freedom of academic pursuit, consistent with empirical evidence by \citet{stern2004scientists}. \citet{aghion2008academic} argue that, in the early stages of research, which is when the commercial value of projects is most uncertain, it is efficient to leave scientists the freedom to decide which questions to pursue, even ``if this entails some probability of the scientists wandering off in other directions.'' Interpreting grants as providing unconditional funding for researchers, this provides a rationale for funding fundamental research via grants.

Fourth, grants may be most suitable when it is difficult to spell out what exactly constitutes a successful research project. 
In some cases, it is roughly clear what success means. 
For example, in 1908, the Wolfskehl Prize was announced for the first person to prove Fermat's Last Theorem. In 1997, the prize was awarded to Andrew Wiles after the scientific community deemed his proof complete.
However, the outcome and the value of an undertaking are often highly uncertain.
In this case, an upfront payment in the form of a grant may constitute a more compelling incentive to tackle a problem than, say, a prize for attaining a nebulously specified goal.

%% file: goal_and_challenges.tex
\section{Challenges in grant allocation}\label{sec:goals_and_challenges}
Despite the merits of grants via-{\`a}-vis patents and prizes, grant allocation is not without its own challenges and pitfalls.
How can grantmakers identify the most meritorious applications? How can grantmakers be certain that, as \citet{gallini2002intellectual} put it, grantees do not ``take the money and run?'' 
The efficiency of the allocation depends on how well grantmakers can screen applications for their merit ex ante and on whether funded researchers work hard to execute their proposals.
Additionally, one would like to achieve all this while economizing on the costs of preparing and reviewing proposals, monitoring funded researchers, etc.

A central premise of modern economics is that institutions are marred by the simultaneous presence of \emph{misaligned interests} and \emph{asymmetric information}.
In the context of a grantmaker deciding whom to fund, it is clear that there are \emph{misaligned interests}: the grantmaker wishes to support the researchers with the ``best'' project, whereas each researcher may primarily care about their own funding.
Asymmetric information is classified into \emph{hidden information} and \emph{hidden action}.
\begin{enumerate}
    \item Hidden information refers to the private knowledge of one party that is relevant for the payoffs or strategic incentives of other parties. For example, on the one hand, researchers may have better information than grantmakers about the intrinsic quality of their proposals, but on the other hand, researchers may be uncertain about what exactly grantmakers hope to find in a proposal.

    \item Hidden actions refer to behavior by one party that others cannot observe or control. For example, grantmakers may be unable to enforce exactly how funded researchers use their grants.
\end{enumerate}
Both forms of asymmetric information seem particularly pertinent in research because research topics and practices are highly specific and require specialist knowledge to evaluate. 
The simultaneous presence of misaligned interests and asymmetric information gives rise to incentive problems.

Prospective and retrospective evaluation are the basic instruments that grantmakers use to mitigate incentive issues.
Prospective evaluation (through referee panels, for example) reduces informational asymmetries prior to the allocation.
Retrospective evaluation (through publication metrics, for example) rewards researchers for their output, thereby aligning researchers' incentives with those of the grantmaker and mitigating the hidden action problem. Retrospective evaluation also has a prospective role, as we explain later.

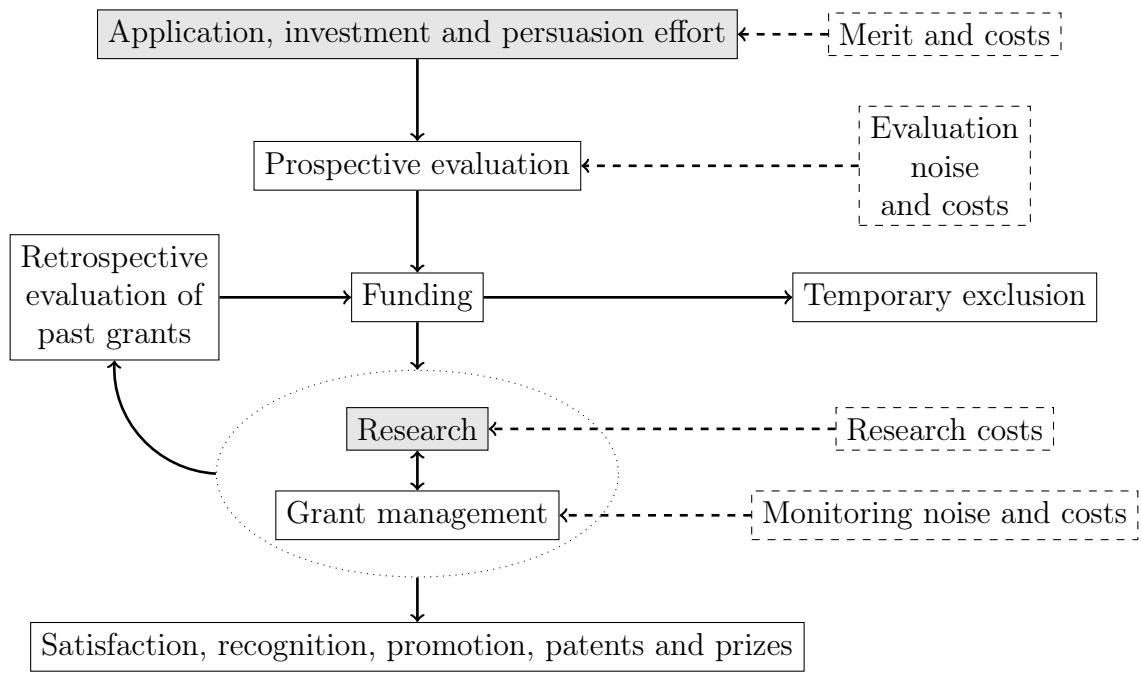
\begin{figure}[ht]
\centering
\begin{tikzpicture}[node distance=1.75cm]
    \node (application) [shade, top color=lightgray!40, bottom color=lightgray!40, draw] {Application, investment and persuasion effort};
    \node (prospective) [draw, rectangle, below of=application, yshift=0cm] {Prospective evaluation};
    \draw [->, line width=1pt] (application) -- (prospective);
    \node (allocation) [draw, rectangle, below of=prospective, yshift=0cm] {Funding};
    \node (retrospective_evaluation) [draw, rectangle, left=1.75cm of allocation, text width=2.5cm, align=center] {Retrospective evaluation of past grants};

    \draw [->, line width=1pt] (prospective) -- (allocation);
    \draw [->, line width=1pt] (retrospective_evaluation) -- (allocation);

    \node (effort) [shade, top color=lightgray!40, bottom color=lightgray!40, draw, below of=allocation] {Research};
    \node (management) [draw, rectangle, below of=effort, yshift=0.6cm] {Grant management};
    \node[fit=(effort) (management), ellipse, draw, dotted, minimum width=5cm, minimum height=2.75cm, inner sep=0pt] (ellipse) {};
    \draw [->, line width=1pt] (allocation) -- (ellipse);
    \draw [<->, line width=1pt] (effort) -- (management);

    \node (results) [draw, rectangle, below of=management, xshift=0cm] {Satisfaction, recognition, promotion, patents and prizes};
    \draw [->, line width=1pt] (ellipse) -- (results);

    \draw [->, line width=1pt, bend right] (ellipse.west) to[in=135, out=45] (retrospective_evaluation.south);

    \node (application_costs) [anchor=left, draw, dashed, rectangle, right=1.2cm of application, yshift=0cm,  align=center] {Merit and costs};
    \draw [->, dashed, line width=1pt] (application_costs) -- (application);
    
    \node (evaluation_noise) [anchor=center, draw, dashed, rectangle, below of=application_costs, yshift=0cm, align=center, text width=2cm] {Evaluation noise and costs};
    \draw [->, dashed, line width=1pt] (evaluation_noise) -- (prospective);
    
    \node (temporary_exclusion) [anchor=left, draw, rectangle, below of=evaluation_noise, yshift=0cm] {Temporary exclusion};  
    \draw [->, line width=1pt] (allocation) -- (temporary_exclusion);
    
    \node (costs_of_research) [anchor=left, draw, dashed, rectangle, below of=temporary_exclusion, yshift=0cm] {Research costs};
    \draw [->, dashed, line width=1pt] (costs_of_research) -- (effort.east);
    
    \node (noisy_performance) [anchor=left, draw, dashed, rectangle, below of=costs_of_research, yshift=0.6cm] {Monitoring noise and costs};   
    \draw [->, dashed, line width=1pt] (noisy_performance) -- (management.east);
\end{tikzpicture}
\caption{A schematic view of grant funding.}
\label{fig:grantfunding}
\end{figure}

\Cref{fig:grantfunding} illustrates schematically the stages of grant funding.
The shaded boxes indicate stages where applicants make choices; the unshaded solid boxes indicate stages where the grantmaker makes a choice; the dashed boxes indicate an impact of outside factors.
When preparing an application, researchers choose how much effort to invest into the intrinsic quality and persuasiveness of their proposals.
When deciding who to fund, the grantmaker both prospectively evaluates the submitted proposals and retrospectively evaluates the researchers' records.
Unsuccessful researchers may be temporarily excluded from reapplying for the same grant.
For successful researchers, the interaction with the grantmaker continues; the researchers decide how to use their funds, while the grantmaker monitors the researchers' output and decides whether to, say, terminate the grant prematurely.
Finally, successful projects reward researchers with accolades in the scientific community and, possibly, monetary rewards in the form of patents and prizes.

In the next section, we analyze in more detail how the forces indicated in \Cref{fig:grantfunding} shape the optimal allocation and use of grants.
In \Cref{sec:application_process}, we study the application process, emphasizing the impact of evaluation noise and application costs.
In \Cref{sec:allocation_rule}, we focus on the allocation rule itself.
Next, in \Cref{sec:postaward_management} we turn to the role of retrospective evaluation and post-award management.
Finally, in \Cref{sec:supply_and_direction} we consider how the grantmaker's choice of what to fund impacts the direction of research.

%% file: grant_design.tex
\section{The application process}\label{sec:application_process}

Grant programs thoroughly evaluate proposals via peer review.
The NIH, for example, is required by law to do so \citep{price2019grants}. Only the applications deemed to be most promising are funded. For an indication of how competitive the process is, consider that the fraction of funded applications for ``Advanced Grants'' at the European Research Council (ERC) is around $10\%$. Preparing an application is a time-consuming and costly activity;\footnote{From surveys of astronomers, \citet{von2015apply} report an average time of 116 hours spent preparing one application. \citet{myers2024tradeoffs}, also using survey data, indicates researchers spend 7 hours per week preparing grant applications.} researchers are more willing to apply when they are more optimistic about obtaining funding.
The costly ordeal of preparing an application thus helps grantmakers screen applicants through self-selection. Conversely, application rates respond to changes in application costs and the evaluation process.

\citet{adda2023grantmaking} analyze the impact of prospective evaluation on the self-selection of applicants. The baseline version of their model can be understood in terms of demand and supply for grants. A more detailed description of their baseline model is in \Cref{box:grantmaking}, further below.

On the demand side, researchers differ in their intrinsic and privately known ``merit.'' To give a concrete example, suppose that merit corresponds to the number of additional papers a researcher can publish if awarded the grant. Researchers enjoy a benefit if funded, but bear application costs.

On the supply side, the grantmaker can only fund a fraction of the applicants and would like to assign grants to the most meritorious applicants. However, the grantmaker only observes a noisy signal about each applicant's merit (via peer review, for example). From the grantmaker's perspective, a higher signal indicates a higher merit of the applicant. Therefore, the grantmaker funds the applicants that obtain a signal above an \textit{acceptance threshold}; the threshold is chosen to exhaust the grantmaker's budget. 

Anticipating this allocation rule, a researcher applies only if they are sufficiently optimistic about being evaluated positively; that is, about generating a signal above the acceptance threshold. Therefore, there is an \textit{application threshold} (distinct from the acceptance threshold) such that only researchers with merit above the application threshold will apply.

Building on this characterization, \citet{adda2023grantmaking} shed light on the effect of changes in the grantmaker's budget and evaluation procedure. First, an increase in the budget of the grantmaker naturally increases the incentive to apply, thereby reducing the average merit of applicants. The effect of a budget increase on the success rate---the ratio of funded applicants to the total number of applicants---is more subtle and depends on the distribution of merit across researchers. On the one hand, a budget increase incentivizes more applications (as just noted), pushing the success rate down. On the other hand, a budget increase allows for more applicants to be funded, pushing the success rate up. Thus, the success rate can either increase or decrease. In particular, it decreases if the elasticity of applications with respect to the budget is greater than one.\footnote{Technically, this case arises when the distribution of researchers' merits follows a distribution with thicker tails than exponential. This assumption is in line with an early observation by \citet{lotka1926frequency} that researchers’ productivity in terms of publications follows a power law.} This appears in line with evidence from the 2009 increase in the budget available for research grants in the US due to Obama's Stimulus Package. Grant applications increased more than the budget, thus resulting in a reduction in the fraction of successful applicants \citep[p.~145]{stephan2012economics}.

A key insight of \citet{adda2023grantmaking} pertains to the impact of evaluation noise on application incentives.
The noise could reflect how carefully evaluators read proposals or whether a funding lottery is used for applicants at the cusp. \citet{adda2023grantmaking} show that a noisier evaluation procedure increases the incentives to apply and thus reduces the self-selection of applicants. Intuitively, as the evaluation becomes noisier, the probability of succeeding in obtaining a grant becomes less responsive to merit, encouraging more low-merit applications (we sketch the logic in more detail below). This result indicates how the details of the allocation rule impact the self-selection of applicants. We provide a more detailed derivation of the result in \Cref{box:grantmaking}.

\begin{modelbox}[label={box:grantmaking}]{Grant allocation and evaluation noise}
    In the following, we set up the single-field model of \citet{adda2023grantmaking} and derive the optimal application behavior and grant allocation rule. Building on the constructed equilibrium, we illustrate how equilibrium outcomes vary with evaluation noise.    
    The grantmaker's budget is denoted by $B$.
    For a researcher, applying comes at a cost $c$, and the value of winning the grant is normalizes to $1$. 
    Let $\theta$ denote the privately known merit of a researcher. From the grantmaker's perspective, merit is a random variable whose cumulative distribution function (CDF) we denote by $G$. The grantmaker observes a noisy signal $x$ about merit. Conditional on merit $\theta$, the signal $x$ is a random variable with CDF $F_{\theta}$, where higher signal realizations indicate higher merit.\footnote{The distribution $F_{\theta}$ is assumed to possess a continuously differentiable and strictly positive density. Moreover, the density is strictly log-supermodular in signal and merit, meaning the monotone likelihood ratio property holds; higher signals indicate high merit.} 
    
    Since signals are informative about merit, the grantmaker optimally sets an acceptance threshold $\hat{x}$ such that all applicants with a signal above $\hat{x}$ are funded. From the perspective of a researcher with merit $\theta$, the winning probability, denoted $W(\theta; \hat{x})$, is thus given by $W(\theta;\hat{x})=1-F_\theta(\hat{x})$. Such a researcher applies if and only if the winning probability exceeds the application costs, meaning $W(\theta;\hat{x}) \geq c$. We denote the application threshold by $\hat{\theta}$. This threshold depends on $\hat{x}$ and coincides with the marginal researcher who is indifferent to applying, meaning $W(\hat{\theta};\hat{x}) = c$ holds. All researchers with merit $\theta \geq \hat{\theta}$ apply.

    In equilibrium, the acceptance threshold $\hat{x}$ is such that the number of awarded grants equals the grantmaker's budget $B$. As each researcher of merit $\theta\geq \hat{\theta}$ wins with probability $W(\theta;\hat{x})$, the threshold $\hat{x}$ solves $B=\int_{\hat{\theta}}^\infty W(\theta;\hat{x})\, dG(\theta)$. 

\begin{center}
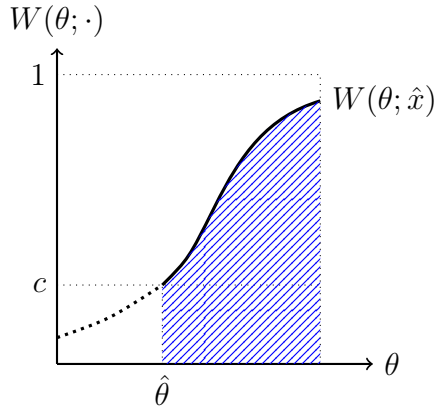

        \begin{tikzpicture}[scale=0.7]
            \draw[thick,->] (0,0) -- (6,0) node[right] {$\theta$};
            \draw[thick,->] (0,0) -- (0,6) node[above]{$W(\theta;\cdot)$};

            \draw[dotted] (0,1.5) node[left] {$c$} -- (5,1.5);
            \draw[dotted] (0,5.5) node[left] {1} -- (5,5.5);

            \draw[very thick] (2,1.5) .. controls (2.5, 2) .. (3,3) .. controls (3.5, 4) and (4, 4.7) .. (5, 5) node[right] {$W(\theta;\hat{x})$};
            \draw[very thick, dotted] (0,0.5) .. controls (1, 0.85) .. (2,1.5) ;

            \fill[pattern=north east lines, pattern color=blue] (2,1.5) .. controls (2.5, 2) .. (3,3) .. controls (3.5, 4) and (4, 4.7) .. (5, 5) -- (5,0) -- (2, 0) -- cycle;

            \draw[dotted] (2,0) node[below] {$\hat{\theta}$} -- (2,1.5);
            \draw[dotted] (5,0) node[below] {} -- (5,5.5);
        \end{tikzpicture}
        \captionof{figure}{Baseline Equilibrium}
        \label{fig:old_eq}
\end{center}

    \Cref{fig:old_eq} depicts this situation for a given threshold $\hat{x}$ assuming that $\theta$ is uniformly distributed. The blue-shaded area in the figure thus represents the mass of awarded grants.

    Consider now a noisier evaluation process, represented by a less accurate signal distribution $\tilde{F}_\theta(x)$.\footnote{Technically, \citet{adda2023grantmaking} consider a decrease with respect to the notion of accuracy introduced by \citet{lehmann1988comparing}.} Noisier evaluation corresponds to less meritocratic acceptance decisions; the probability of winning the grant is \emph{less steeply} increasing in merit. Pictorially, the function $W$ changes to a flatter function $\tilde{W}$. Suppose for a moment that the new acceptance threshold $\tilde{x}$ is such that the same merit type $\hat{\theta}$ is indifferent; that is, $\tilde{W}(\hat{\theta};\tilde{x})=1-\tilde{F}_{\hat{\theta}}(\tilde{x})=c$. The left panel of \Cref{fig:new_eq} depicts this scenario. Since the new winning probability is flatter, the grantmaker does not exhaust the budget. The red-shaded area in the left panel of \Cref{fig:new_eq} depicts the unspent budget. 

\begin{center}
        \begin{tikzpicture}[scale=0.6]
            \draw[thick,->] (0,0) -- (6,0) node[right] {$\theta$};
            \draw[thick,->] (0,0) -- (0,6) node[above]{$W(\theta;\cdot)$};

            \draw[dotted] (0,1.5) node[left] {$c$} -- (5,1.5);
            \draw[dotted] (0,5.5) node[left] {1} -- (5,5.5);

            \draw[thick, dashed] (2,1.5) .. controls (2.5, 2) .. (3,3) .. controls (3.5, 4) and (4, 4.7) .. (5, 5) node[right] {$W(\theta;\hat{x})$};
            \draw[thick, dotted] (0,0.5) .. controls (1, 0.85) .. (2,1.5) ;

            \draw[very thick, black] (2,1.5) .. controls (3,1.7) and (3.8,3.4) .. (5,3.5) node[right] {$\tilde{W}(\theta;\tilde{x})$};
            \draw[very thick, dotted, black] (0,1) .. controls (1, 1.1) .. (2,1.5) ;

            \fill[pattern=north west lines, pattern color=darkred] (2,1.5) .. controls (2.5, 2) .. (3,3) .. controls (3.5, 4) and (4, 4.7) .. (5, 5) -- (5,3.5) .. controls (3.8,3.4) and (3,1.7) .. (2,1.5) -- cycle;
            
            \fill[pattern=north east lines, pattern color=blue] (0,0) -- (5,0) -- (5,3.5) .. controls (3.8,3.4) and (3,1.7) .. (2,1.5) -- (2,0) -- cycle;

            \draw[dotted] (2,0) node[below] {$\hat{\theta}$} -- (2,1.5);
            \draw[dotted] (5,0) node[below] {} -- (5,5.5);
        \end{tikzpicture}
\hspace{1cm}
        \begin{tikzpicture}[scale=0.6]
            \draw[thick,->] (0,0) -- (6,0) node[right] {$\theta$};
            \draw[thick,->] (0,0) -- (0,6) node[above]{$W(\theta;\cdot)$};

            \draw[dotted] (0,1.5) node[left] {$c$} -- (5,1.5);
            \draw[dotted] (0,5.5) node[left] {1} -- (5,5.5);

            \draw[dashed, thick] (2,1.5) .. controls (2.5, 2) .. (3,3) .. controls (3.5, 4) and (4, 4.7) .. (5, 5) node[right] {$W(\theta;\hat{x})$};
            \draw[thick, dotted] (0,0.5) .. controls (1, 0.85) .. (2,1.5) ;

            \draw[dashed, thick] (2,1.5) .. controls (3,1.7) and (3.8,3.4) .. (5,3.5) node[right, yshift=-0.2cm] {$\tilde{W}(\theta;\tilde{x})$};
            \draw[dotted, thick] (0,1.1) .. controls (1, 1.2) .. (2,1.5) ;

            \draw[very thick, blue] (1,1.5) .. controls (2, 1.8) .. (3,2.7) .. controls (4, 3.8) and (4.5, 3.9) .. (5, 4) node[right] {$\tilde{W}(\theta;\tilde{x}^\ast)$};
            \draw[very thick, blue, dotted] (0,1.39) .. controls (0.75, 1.43) .. (1,1.5) ;

            \draw[dotted] (2,0) node[below] {$\hat{\theta}$} -- (2,1.5);
            \draw[dotted] (5,0) node[below] {} -- (5,5.5);

            \fill[pattern=north east lines, pattern color=darkgreen] (1,1.5) .. controls (2, 1.8) .. (3,2.7) .. controls (2.4, 2) .. (2,1.5) -- cycle;
            \fill[pattern=north east lines, pattern color=darkgreen] (1, 0) -- (1, 1.5) -- (2, 1.5) -- (2, 0);

            \fill[pattern=north west lines, pattern color=darkred] (5, 4) .. controls (4.5, 3.9) and (4, 4.2) .. (2.5,2) .. controls (3.5, 4) and (4, 4.7) .. (5, 5) -- cycle;

            \draw[dotted] (1,0) node[below] {$\tilde{\theta}$} -- (1,1.5);
            \node at (1.5, -0.75) {\footnotesize{$\leftarrow$}};
        \end{tikzpicture}
\end{center}
        \vspace{-0.5cm}
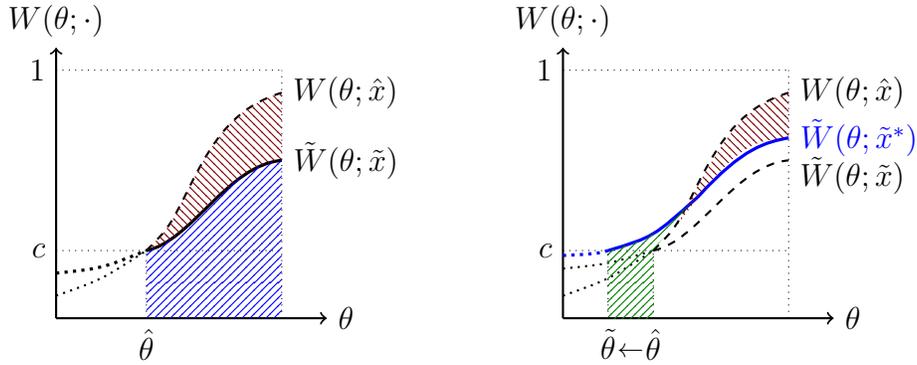
\captionof{figure}{Equilibrium with Noisier Evaluation}
        \label{fig:new_eq}

    To exhaust the budget, the grantmaker has to encourage more applications. That is, the grantmaker decreases the acceptance threshold to a point $\tilde{x}^{\ast}$, thereby raising the winning probability $\tilde{W}$ for all merit types. The new marginal researcher lies at a point $\tilde{\theta}$ below $\hat{\theta}$.
    In the right panel of \Cref{fig:new_eq}, the green area indicates the volume of additional budget that is spent by decreasing the acceptance threshold to $\tilde{x}^{\ast}$; the red and green areas are equal.
    Thus, in the new equilibrium under noisier evaluation, there are more applications and, therefore, the average merit of applicants is lower.
\end{modelbox}

The apportionment of budget across different fields is more delicate yet. Funding organizations grapple with the incentives of panel members to favor research in their own fields. Panel members have an incentive to inflate the scores of applications in their field in an attempt to secure a larger fraction of the total budget available for distribution to all panels.
To counteract this incentive, some of the world's largest research funding organizations have been allocating funds depending on the relative ranking of applications received in each panel, thus apportioning the overall budget according to a mechanical formula. For example, the payline system adopted in 1988 by NIH institutes equalizes the success rate across panels. Since its inception in 2007, the ERC adopted an equivalent scheme by splitting its budget in proportion to the funds requested by applicants in each panel, thus automatically equalizing the success rate across panels. 

At first glance, proportional apportionment seems fair and balanced. In the framework of \citet{adda2023grantmaking}, the system indeed performs well if fields are relatively similar in terms of the noise in the evaluation signal. However, if fields are heterogeneous in terms of noise, the performance deteriorates. Noisier fields attract more applications (as argued above), which, under proportional apportionment, leads to a proportional increase in the budget; this budget increase, in turn, induces a further increase in applications. To see how adverse this effect can be, consider as an extreme the case of a field with perfect evaluation (meaning all applicants correctly anticipate whether they will succeed). Applicants who anticipate failing do not apply, and thereby reduce the funding for other applicants from the same field. Whenever the success rate is less than 100\%, the process will continue until, in equilibrium, no applications at all will be submitted from this field.

Beyond this stark illustration, \citet{adda2023grantmaking} show that under a general class of apportionment rules, including the proportional rule, a reduction of noise in a field leads to a reduction in applications in that field and an increase in applications in all other fields. 

\citet{adda2023grantmaking} also present a broad empirical confirmation of their predictions by exploiting a natural experiment. A 2014 reform changed the proportional budget allocation rule at the ERC. After the reform, a panel's budget depended not only on applications in its own domain (e.g., life sciences) but also on the number of applications in the other domains (e.g., social sciences and humanities).

\begin{figure}[t!]
\centering
\includegraphics[width=0.9\textwidth]{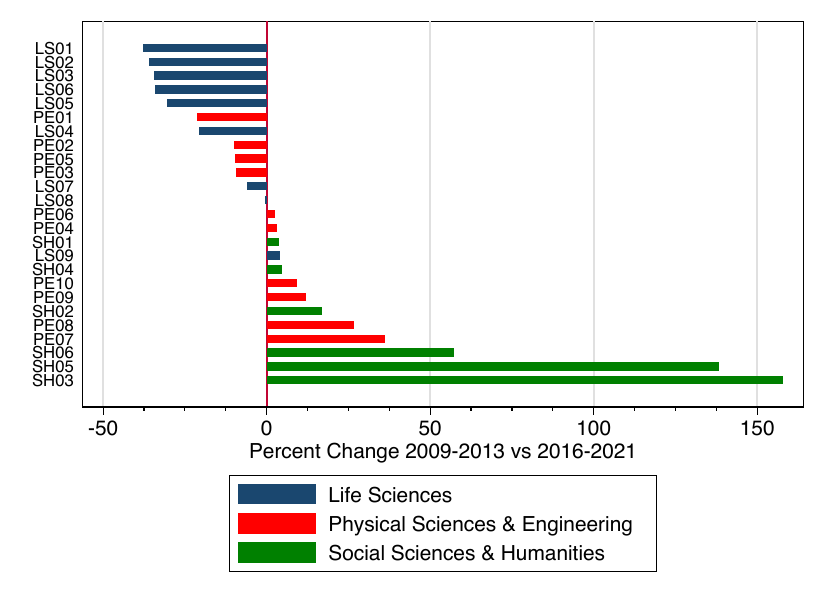}
\caption{The percent change in the budget of each ERC panel after the reform.}
\label{fig:grantmaking_empirical:budget}
\end{figure}

\Cref{fig:grantmaking_empirical:budget} reports the percentage change in budget for each ERC panel after the reform in the budget allocation rule, compared to pre-reform levels. Panels in the social sciences and more applied areas of physical sciences and engineering received increased budgets, primarily at the expense of panels in the life sciences and more foundational areas of physical sciences and engineering. 

\begin{figure}[t!]
\centering
\includegraphics[width=0.9\textwidth]{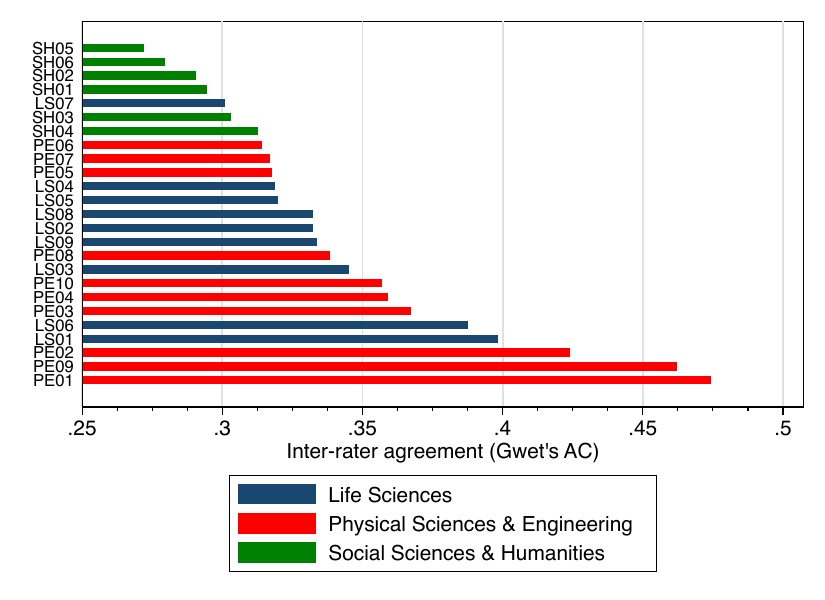}
\caption{The inter-rater agreement by panel in evaluating RCN funding applications.}
\label{fig:grantmaking_empirical:disagreement}
\end{figure}

\Cref{fig:grantmaking_empirical:disagreement} reports a measure of consensus (Gwet’s inter-rater agreement) among reviewers for grant proposals submitted to the Research Council of Norway, with abstracts similar to those submitted to the corresponding ERC panels. In line with theory, \citet{adda2023grantmaking} find that on average panels with less agreement, and thus more evaluation noise, gained budget after the reform.

In terms of efficiency, it is undesirable that noisier fields attract more applications and more funding.
After all, in noisy fields, it is difficult to allocate efficiently, and hence, all else equal, these fields should be endowed with little funding. An important challenge is to identify rules that are both efficient and politically feasible.

A second important challenge is to reduce the application and evaluation costs inherent to this system.
We briefly discuss \emph{temporary exclusion} with this challenge in mind.
By temporary exclusion, we mean that unsuccessful applicants are debarred from reapplying for some period of time. At the ERC, for example, applicants to Starting or Consolidator Grants who are rejected in the early stages of the evaluation process may not reapply to the same grant for two subsequent years \citep[p. 9]{erc2023information}.\footnote{See \url{https://erc.europa.eu/apply-grant/starting-grant}.}  The aim of this restriction is ``to allow unsuccessful Principal Investigators the time necessary to develop a stronger proposal.''

\citet{azrieli2024temporary} studies temporary exclusion in a dynamic model similar to the model of \citet{adda2023grantmaking}. 
The threat of temporary exclusion deters researchers from applying if they know that their current project is weak, but they expect to have a stronger one in the future.
This basic logic suggests temporary exclusion indeed reduces excessive applications.
The full picture is more nuanced because of a countervailing force: if one's competitors are less likely to apply, then one's own incentive to apply increases. \citet{azrieli2024temporary} shows that temporary exclusion can at once increase the overall welfare of researchers and decrease the number of applications. In particular, this result applies when the costs of applying are small relative to the gains from obtaining a grant.

\paragraph{Further reading} \citet{pereyra2023optimal} study a related setting where, roughly speaking, the grantmaker allocates grants of heterogeneous quality (for example, they may in the funding duration). As in \citet{adda2023grantmaking}, the grantmaker only observes a noisy signal about researchers' privately known types, but researchers bear no application costs. To maximize welfare, the grantmaker lets researchers self-select into one of multiple tracks. In ``tougher'' tracks, researchers need more positive evaluations to be awarded a high-quality grant.\footnote{\citet{pereyra2023optimal} do not explicitly cast their model as one of grant allocation. Any faults with the interpretation of their model in the context of grant allocation are due to us. Another important difference to \citet{adda2023grantmaking} is that \citet{pereyra2023optimal} cast the problem as one of mechanism design. In particular, the grantmaker has commitment power. We discuss mechanism design in more detail in \Cref{sec:allocation_rule}.}

\section{The allocation rule}\label{sec:allocation_rule}

In the previous section, we focused on the application process, taking a stylized view on how the grantmaker allocates grants to applicants.
We now zoom in on optimal allocation rules through the lens of \emph{mechanism design}.

Mechanism design is a subfield of economic theory that studies how to design institutions---\emph{mechanisms}---to achieve a pre-specified goal. 
Typically, this goal will not be perfectly achievable.
For example, a grantmaker may wish to award grants to the most fund-worthy applicants while economizing on the time spent reviewing proposals. 
The applicants, who know the nuts and bolts of their proposals, have private information about their fund-worthiness that should guide the efficient allocation. 
However, if the grantmaker would completely save on evaluation costs by never triggering a review, thereby solely relying on what the applicants claim in their proposals, then the applicants would have incentives to overstate their fund-worthiness. 
Hence, we do not expect that the grantmaker can achieve the first-best outcome of funding the best proposals at zero cost. 
This impossibility is a consequence of the simultaneous presence of misaligned interests and asymmetric information (and the fact that there are not enough resources for everyone).

When the first-best is out of reach, mechanism design can shed light on achievable \emph{second}-best outcomes: how does a mechanism optimally balance allocative efficiency with evaluation costs? For example, we shall see that optimal mechanisms feature some deliberate randomization in some but not in all environments. Thus, here, the analysis suggests a practically relevant feature of allocation rules and illuminates which properties of the environment justify this feature from an efficiency perspective. More conceptually, the analysis provides guidance on how grantmakers can benefit from combining certain instruments (such as a costly review process) with self-reports by the applicants.

Notice that we take the objective of the grantmaker as given; the question is how to implement it. We will be silent on what the grantmaker's notion of fund-worthiness should be, or on how the grantmaker's objective should weigh fund-worthiness against the time burden imposed on the reviewers.

Before delving into the details, we emphasize that, unless mentioned otherwise, the models discussed below are cast as general allocation problems. In particular, we do not claim that the papers from which we draw are focused on the specific problem of allocating grants. Any faults with the interpretation of the papers' results in the context of grant allocation lie with us.

\subsection{Costly prospective evaluation}\label{sec:design_aspects:costly_evaluation}

How can a grantmaker fund the most promising research while economizing on evaluation costs? We study this question using the model of \citet{benporath2014optimal}.

Consider a grantmaker who has one grant to allocate among a group of applicants.
Each applicant wants the grant for themselves and privately knows their merit.
For the sake of concreteness, suppose each applicant has some estimate about the number of additional publications they could produce if funded. The grantmaker wishes to maximize this number and has no other use for the funds.
If the grantmaker were to ask each applicant to self-report their estimate and then fund the top applicant, the applicants would have an incentive to exaggerate.
The grantmaker has an additional instrument at their disposal: they can \emph{verify} individual applicants at a cost. Verifying an applicant reveals the applicant's merit. We interpret the act of verifying as conducting an in-depth review by external evaluators; verification costs represent the time burden this review imposes on the evaluators, capturing the idea that the grantmaker wants to economize on evaluation. 
We assume here that the applicants are ex-ante symmetric (with respect to the distribution of their merit levels and the costs of verifying them); \citet{benporath2014optimal} also cover the asymmetric case.

Let us elaborate on the interpretation \citep[p. 3803]{benporath2014optimal}. What really matters is that verification reveals everything the applicant knows; that is, if the applicant's estimate for the number of additional publications is pinned down by a set of objective facts about the applicant's research that are ex-ante only known to the applicant, then verification reveals those facts. There could well be residual uncertainty about the number of additional publications, and neither the applicant nor the principal can presently resolve this uncertainty.

In this environment, a mechanism specifies which reports the applicants can make to the grantmaker, which applicants are verified depending on those reports, and how the grant is eventually allocated depending on the reports and verification outcomes.
To illustrate, here are three examples of mechanisms. First, the grantmaker could randomly select a winner without requesting or verifying reports. Second, the grantmaker could outright verify all applicants and then allocate the grant to the most meritorious one. Third, the grantmaker could approach applicants sequentially, asking for self-declarations about merit, possibly verifying some of them, and stopping when an applicant seems ``good enough.''  
Given a mechanism, applicants strategically choose which reports to make to maximize the probability of receiving the grant, forming conjectures about what their competitors might do. The scope of possible mechanisms and outcomes is thus quite complex.

Which mechanisms maximize the expected merit of the funded applicant net of verification costs? 
\citet[Theorem 1]{benporath2014optimal} show that there is an optimal mechanism of the following form:
The grantmaker announces a merit threshold and asks all applicants to self-report their merit. If all reports are below the threshold, the grant is allocated uniformly at random. If, instead, at least one report is above the threshold, then the grantmaker verifies the highest report and, if the report is verified to be true, awards the grant to the corresponding applicant.
If an applicant is verified and found to have misreported their merit, then that applicant is \emph{not} allocated the grant.\footnote{In this case, as long as the misreporting applicant is denied the grant, there is a lot of flexibility regarding how to proceed with the allocation. The reason for this flexibility is that in the described mechanism, the applicants expect that everyone will report truthfully, and therefore, they completely disregard the event in which one of the other applicants is verified to have misreported. One way of proceeding would be to allocate the grant randomly among the remaining agents or not to allocate at all. Another perhaps more reasonable way is to sequentially verify the remaining highest reports until someone is verified to have been truthful, and then award that applicant; in case everyone is verified to have been untruthful, no one is awarded.} Given these allocation and verification rules, all applicants find it in their best interest to report truthfully; indeed, when untruthfully claiming a merit above the threshold, an applicant anticipates they will be detected. In \Cref{box:costly_evaluation}, we illustrate the one-applicant version of this model and show how to derive the optimal allocation mechanism.

The described optimal mechanism allocates the grant through a lottery when no report is sufficiently promising. This feature speaks to the growing interest in lotteries among real-world funders \citep{heyard2022rethinking,bendiscioli2022handbook}. A key argument put forth by \citet{fang2016research} in favor of lotteries is that identifying the best proposals is extremely costly. This concern is roughly reflected in the optimal mechanism. The grantmaker \emph{could} verify all reports and identify the best applicant. However, the grantmaker chooses not to do so, saving on verification costs by randomly allocating the grant when all reported merits are below the threshold. 
That said, the mechanism identifies and allocates to the best applicant whenever the highest merit is above the threshold. 
Further, the use of a lottery does depend on the assumption that the applicants are ex-ante symmetric.\footnote{In the asymmetric case, there is an optimal mechanism that (for almost all realizations) \emph{deterministically} allocates to a so-called ``favored agent'' if no self-report passes the threshold \citep[Theorem 1]{benporath2014optimal}.}

\citet[Section IV]{benporath2014optimal} show that for higher verification costs, the optimal mechanism uses a higher threshold; so, there are fewer in-depth reviews, and allocation is more frequently via a lottery.
They also show that the threshold increases in response to shifts of the distribution of merit in the sense of first-order or second-order stochastic dominance; that is, roughly speaking, if applicants become more meritorious or less heterogeneous, then the threshold increases, meaning the allocation is more frequently via a lottery.

\begin{modelbox}[label=box:costly_evaluation]{Optimal allocation with costly evaluation}
    Let us consider a single-applicant version of the problem in \citet{benporath2014optimal}.
    The grantmaker's decision is now between funding the applicant and keeping the funds.
    The applicant only cares about the funding probability. 
    The grantmaker's payoff from keeping the funds is normalized to $0$. Let $\theta$ denote the applicant's privately known merit type. The grantmaker's payoff from funding the applicant equals $\theta$, and $\theta$ may be greater or less than $0$.
    From the grantmaker's perspective, the merit type $\theta$ is a random variable whose cumulative distribution function we denote by $F$. The verification costs are denoted by $c$. 
    
    For optimality, nothing is lost by focusing on mechanisms of the following form.\footnote{This step combines a version of the Revelation Principle with optimality considerations.} For each merit type $\hat{\theta}$, the mechanism specifies two probabilities $v(\hat{\theta})$ and $q(\hat{\theta})$, and the mechanism plays out as follows. First, the applicant is asked to report their merit. Given the applicant's report $\hat{\theta}$, the applicant is verified with probability $v(\hat{\theta})$; if the applicant is verified, the applicant is funded for sure if the report was truthful ($\theta = \hat{\theta}$), and with probability $0$ if untruthful; if the applicant's report is not verified, the applicant is funded with probability $q(\hat{\theta})$.
    Moreover, in the mechanism, the applicant always finds it in their best interest to report truthfully; that is, for all $\theta$ and $\hat{\theta}$,
    \begin{align}\label{eq:BDL:IC}
        \tag{IC}
        v(\theta)\cdot 1 + (1 - v(\theta))\cdot q(\theta) \geq (1 - v(\hat{\theta}))\cdot q(\hat{\theta})
    \end{align}
    The left side of this inequality is the overall winning probability for merit-type $\theta$ when reporting truthfully; the right side when misreporting $\hat{\theta}$. This inequality is known as \emph{incentive compatibility (IC) for type $\theta$.}
    
    It is convenient to define the overall winning probability as $p(\theta) = v(\theta)\cdot 1 + (1 - v(\theta))\cdot q(\theta)$. Thus \eqref{eq:BDL:IC} can be written as $p(\theta) \geq p(\hat{\theta}) - v(\hat{\theta})$. Since this inequality holds for all $\theta$ and $\hat{\theta}$, it also holds if we flip their roles, yielding $p(\hat{\theta}) \geq p(\theta) - v(\theta)$. We can now rearrange to obtain
    \begin{align}\label{eq:BDL:ICast}
        \tag{IC*}
        v(\theta) \geq p(\theta) - p(\hat{\theta}).
    \end{align}
    This inequality says that the probability $v(\theta)$ of verifying $\theta$ must be sufficiently high to deter $\hat{\theta}$ from misreporting to $\theta$.
    
    The grantmaker's expected payoff equals
    \begin{align*}
        \mathbb{E}_{\theta\sim F}[p(\theta)\cdot \theta + (1 - p(\theta))\cdot 0 - v(\theta) \cdot c].
    \end{align*}
    We want to maximize the grantmaker's payoff subject to \eqref{eq:BDL:ICast}. To see how this is done, let $\ubar{p}$ denote the winning probability of the worst-off merit types; that is, $\ubar{p} = \inf_{\theta\in\mathbb{R}} p(\theta)$. The truth-telling constraint \eqref{eq:BDL:ICast} is hardest to satisfy for the worst-off merit types; that is, \eqref{eq:BDL:ICast} holds if and only if $v(\theta) \geq p(\theta) - \ubar{p}$ holds for all $\theta$. The grantmaker wishes to save on verification costs and therefore sets $v(\theta)$ as low as possible, meaning $v(\theta) = p(\theta) - \ubar{p}$.
    Plugging this equation into the grantmaker's expected payoff and rearranging yields
    \begin{align*}
        \mathbb{E}_{\theta\sim F}[(p(\theta) - \ubar{p})\cdot(\theta - c) + \ubar{p} \cdot \theta].
    \end{align*}
    This expected payoff has the following interpretation: when the grantmaker wishes to raise $p(\theta)$ beyond the value $\ubar{p}$, the grantmaker must simultaneously raise the verification probability $v(\theta)$ since, else, the worst-off merit types would misreport. Thus, the gain from raising $p(\theta)$ effectively equals the \emph{net} merit type  $\theta - c$.
    It follows that, optimally, $p(\theta)$ equals $1$ when the net merit type is positive ($\theta - c \geq 0$) and equals $0$ when the net merit type is negative.
    
    The optimal value of $\ubar{p}$ depends on the parameters.
    For example, setting $\ubar{p} = 0$ is optimal if the prior mean $\mathbb{E}_{\theta\sim F}[\theta]$ is less than $0$ (which was the normalized value of keeping the funds), capturing a scenario where the grantmaker is a priori unwilling to fund the applicant.\footnote{Let us show this. Suppose $p(\theta)$ equals $1$ if $\theta - c \geq 0$, and equals $\ubar{p}$ otherwise. We can now write the grantmaker's expected payoff as
    \begin{align*}
        (1 - \ubar{p})\mathbb{E}_{\theta\sim F}[ \max(0, \theta - c)] + \ubar{p}\mathbb{E}_{\theta\sim F}[\theta].
    \end{align*}
    In this expression, the first expectation is positive. The second is negative (by assumption). Thus the grantmaker's expected payoff is decreasing in $\ubar{p}$; optimally, $\ubar{p}$ equals $0$.}
    
    When $\ubar{p}$ equals $0$, the optimal mechanism can be described as follows. The applicant reports truthfully. Reports below $0$ are not verified and do not lead to funding. Reports $\hat{\theta}$ above $c$ are verified with certainty and lead to funding if and only if the report is truthful.
\end{modelbox}

In practice, grantmakers may not only rely on the outcome of an in-depth review; they also have access to a relatively cheap but noisy signal, perhaps from a preliminary screening of the applicants. \citet{kattwinkel2023costless} study how to optimally combine such a noisy signal with self-reports and verification in a setting with a single applicant (thus, the decision is between funding the applicant and keeping the funds). 
Kattwinkel and Knoepfle show that, optimally, the grantmaker reveals the results of the preliminary review to the applicant, and then allows the applicant to submit an appeal to be verified. The applicant is funded if the preliminary review is sufficiently positive, or if the applicant is verified to have sufficiently high merit upon submitting an appeal. 
An intriguing feature of this mechanism is that the grantmaker does not benefit from obscuring the result of the preliminary review; the optimal transparent procedure does not sacrifice allocative efficiency.
This feature is surprising since, in theory, the principal could try to cross-check the agent's self-report with the preliminary review to deliver stronger incentives for reporting truthfully, potentially letting the grantmaker save on verification costs.

\paragraph{Further reading} \citet{ben2019mechanisms,ben2023sequential} consider, among other things, nearby settings where the grantmaker cannot verify the applicants' claims; instead, the applicants themselves can provide hard evidence about their merit but must be incentivized to do so.
\citet{khalfan2023optimal} studies a problem where applicants only have a noisy signal about their merit and hence do not know what an in-depth review will reveal. This detail complicates the grantmaker's problem as applicants can now feign ignorance if a review reveals low merit.
\citet{epitropou2019optimal} consider a problem where the applicants arrive sequentially. \citet{li2023dynamics} consider a repeated interaction with a single applicant.

\subsection{Costly ordeals}

As highlighted by \citet{adda2023grantmaking}, costly ordeals can help grantmakers screen applicants and improve allocative efficiency.
But, of course, an ordeal itself is a wasteful activity, raising the question of optimally balancing costly ordeals and allocative efficiency from a welfare perspective.
We explore this question in a different framework based on \citet{condorelli2012money}.
As we highlight below, depending on the parameters of the environment, it may be optimal to allocate efficiently while imposing highly costly ordeals, but it may also be optimal to simply allocate grants at random while imposing no costly ordeals.

In the model, a grantmaker has a fixed set of grants to be distributed among a fixed set of researchers.
The grants potentially differ in their desirability (for example, in the amount of funds or the funding duration).
There are not enough grants for everyone.
Researchers differ in their privately known valuations for funding (we interpret this valuation momentarily).
Prior to allocating the grants, the grantmaker asks researchers to report their valuations and then imposes costly ordeals on the researchers.
An ordeal may consist of filling out forms whose length is commensurate with one's claimed valuation but with no informational value on top of one's report.
Researchers with higher valuations will have a higher willingness to bear any given ordeal.
Of course, researchers can walk away at any time to avoid the ordeal and forgo the chance at funding.
The grantmaker seeks to maximize the expected valuation across funded researchers minus the costs of the imposed ordeals, potentially assigning different welfare weights to different researchers.

We interpret a researcher's valuation as capturing, for example, the privately known scientific merit of the researcher's project as well as how desperate the researcher needs funds for purchasing laboratory equipment.
In this interpretation, the researcher's willingness to endure an ordeal is increasing in their merit and funding need.

\citet{condorelli2012money} characterizes the following optimal mechanism. The grantmaker computes for each researcher and reported valuation a \textit{priority} (explained momentarily) and then assigns the most desirable grants assortatively to the researchers with the highest priorities (until all grants are assigned). The ordeals are chosen to elicit truthful claims. Researchers with higher claimed valuations bear more costly ordeals to deter exaggerated claims.

A researcher's priority is increasing in their reported valuation, but not necessarily strictly increasing.
That is, researchers of different valuations may be assigned the same priority.
As described by Condorelli, the exact relationship between valuation and priority depends on the prior distribution of valuations.
We highlight two cases.

First, suppose each researcher's valuation distribution has a \textit{weakly increasing} hazard rate.
In this case, each researcher's priority is constant in the reported valuation; a researcher's priority depends only on their expected valuation and their welfare weight in the grantmaker's objective function.
The described optimal mechanism thus assigns the grants \emph{without} eliciting reports or imposing ordeals. If researchers are symmetric (in terms of valuation distributions and welfare weights), the mechanism simplifies to a lottery.

Second, suppose each researcher's valuation distribution has a \textit{strictly decreasing} hazard rate. In this case, each researcher's priority is strictly increasing in their valuation. If researchers are symmetric, the described optimal mechanism thus allocates the best grants to the highest valuation researchers. The mechanism maximizes allocative efficiency but relies on costly ordeals to elicit the researchers' private information.

\paragraph{Further reading} See work by \citet{chakravarty2013optimal,yoon2011optimal,hartline2008optimal} for similar analyses. \citet{dworczak2023equity} provides a sufficient condition for optimal mechanisms to involve costly ordeals. Dworczak's model is mathematically similar to Condorelli's, but the interpretations differ somewhat; see \citet[Section 2.1]{dworczak2023equity} for elaboration.

\subsection{Peer selection}

Next, we discuss a setting in which the grant applicants and the reviewers are the same group of people. 

On an abstract level, this setting fits the problem of distributing resources within the scientific community. There is a finite amount of resources to allocate. To allocate these efficiently, grantmakers would benefit from knowing how each scientist evaluates their peers.
Of course, there is no single grantmaker who allocates all resources, and not all scientists simultaneously apply for a grant, but the model serves as a useful abstraction to shed light on an economic tension (explained momentarily).

On a more concrete level, it has been suggested to \emph{require} applicants to evaluate their competitors' proposals. \citet{merrifield2009telescope} suggest such a procedure for allocating telescope time to alleviate the burden that a conventional process places on external reviewers. A 2014 NSF pilot later used such a procedure to allocate grants; the Gemini Observatory to allocate time on its telescope in Hawaii \citep{mervis2016radicalchange}. The 2016 Neural Information Processing Systems conference, a top-tier machine learning conference, asked authors to volunteer as reviewers to handle the enormous number of submissions \citep{shah2018design}.

Here, we focus on one particular issue that appears relevant when applicants review their competitors: dishonestly exaggerating one's merits and others' faults. See, for example, the laboratory experiment of \citet{balietti2016peer} or the field experiment of \citet{hussam2022targeting} for evidence that individuals indeed misreport when evaluating their peers (albeit not in the context of scientific grant funding).

We consider a stylized setting, borrowing from work by \citet{alon2011sum}.
Suppose each applicant has an honest private opinion about who they would nominate as the most deserving recipient of the grant. The grantmaker wishes to allocate to the applicant with the most nominations. At the same time, the grantmaker wishes to use a mechanism under which no applicant can influence their own chance of winning by misreporting their nomination. In the literature, such a rule is called \emph{impartial}.\footnote{Some papers instead use the terms \emph{strategyproofness} or \emph{dominant-strategy incentive-compatibility} instead of \emph{impartiality}. These notions all coincide under the assumption that each applicant seeks to maximize their own chance of winning and is otherwise indifferent to who wins. \label{footnote:impartial_synonyms}}

The key economic tension in this problem stems from the grantmaker's desire to elicit a relative ranking of the applicants. Applicants improve their rank not only by appraising themselves but also by diminishing their peers. Even if self-nominations are prohibited, an applicant may be able to improve their rank by claiming that all others are unworthy of the grant.

Here is one example of an impartial mechanism. First, before eliciting any nominations, the grantmaker randomly partitions the pool of applicants into two groups---``evaluators'' and ``candidates''. Second, the grantmaker counts only the nominations of the evaluators to pick a winner among the candidates.

Under this partition mechanism, evaluators know that they are excluded from winning (and hence cannot improve their chances by misreporting), and candidates know that their reports are not counted. Instead of randomly splitting applicants into two groups to award one grant, one could also split the grant into two parts and have the evaluators for the first part be the candidates for the second part and vice versa.

The partition mechanism is intuitive and takes to heart a central tenet of peer review---evaluators should have no personal stake in the funding decision---but the mechanism is not flawless.
The partition mechanism yields a bad outcome if all applicants with few nominations are randomly assigned the role of candidate.

\citet{fischer2014optimal} show that the grantmaker can improve on the partition mechanism by splitting the applicants into more than two groups (the procedure for aggregating their nominations becomes more intricate, though). Fischer and Klimm propose a mechanism that guarantees to select an applicant who receives at least half as many approvals as the applicant with the most approvals. 
The fraction of one-half is the best theoretical guarantee \citep{alon2011sum,fischer2014optimal}. \citet{niemeyer2023simpleallocation}, among other things, provide conditions under which the partition mechanism is approximately optimal among all impartial mechanisms when there are many applicants, albeit under a different notion of optimality.\footnote{\citet{alon2011sum,fischer2014optimal} evaluate a rule using a worst-case criterion: across all possible nomination realizations, in the worst case, what is the ratio between the nominations received by the chosen applicant and the most nominations received by any applicant? \citet{niemeyer2023simpleallocation} instead consider the number of nominations of the chosen applicant, taking expectations with respect to the grantmaker's belief about the realized nominations.}

\paragraph{Further reading} A large number of papers build on the framework of \citet{alon2011sum}. For example, \citet{kurokawa2015impartial} impose the additional constraint that each applicant can only review a limited number of other proposals. \citet{holzman2013impartial,de2008impartial,mackenzie2015symmetry}, among others, instead study which other desirable properties are compatible with impartiality. \citet{olckers2023manipulation} survey the literature on impartial mechanisms. In related settings, \citet{kattwinkel2022mechanisms,bloch2023selecting,kattwinkel2020allocation} consider mechanisms that are not necessarily impartial but which nevertheless provide the applicants with incentives for truthful reports;\footnote{In technical terms, these authors consider \emph{Bayesian incentive-compatible} mechanisms. As noted in \cref{footnote:impartial_synonyms}, impartiality coincides with the stronger notion of dominant-strategy incentive-compatibility.} for example, they allow for mechanisms where the grantmaker cross-checks reports across applicants or with external reviewers.

\subsection{Incentivizing investment}
\label{sec:design_aspects:productive_investment}
Up to this point, we treated each applicant's merit as exogenous; applicants do not control the quality of their proposals. However, we would think that in practice researchers adjust their proposals in response to, say, the fierceness of their competition or the grantmaker's evaluation criteria.

In this section, we discuss how the grantmaker's mechanism incentivizes productive investment, based on work by \citet{augias2023nonmarket}.
The headline insight is that the grantmaker does \emph{not} benefit from allocating randomly. Moreover, the optimal evaluation is noisy for low-merit proposals but exact for high-merit ones.

In the model, a unit mass of grants can be assigned to a unit mass of applicants. 
Applicants are heterogeneous with respect to their initial merit level.
Crucially, applicants can invest in their merit, incurring a cost that increases in the distance between the initial and final merit.
Applicants care about the probability of receiving funding minus the investment costs. 
The grantmaker observes each applicant's final merit (but neither initial merit nor investment) and then decides whom to give a grant. 
In the baseline model, the grantmaker has enough resources to fund everyone, but would only like to fund those whose final merit exceeds some threshold; let us call this the grantmaker's preferred threshold.

\citet[Theorem 1]{augias2023nonmarket} provide a condition on the distribution of initial merit under which a deterministic threshold mechanism is optimal.\footnote{Interestingly, a deterministic threshold mechanism is also optimal (under the condition on the merit distribution) if the grantmaker is concerned with welfare \citep[Proposition 5]{augias2023nonmarket}. Here, welfare means the final merit of the funded applicants, plus the private payoffs of the funded applicants from winning, minus total investment costs. However, the value of the threshold changes relative to the case where the grantmaker only cares about final merit.} This mechanism deterministically funds the applicants whose final merit exceeds the mechanism's acceptance threshold; all others are rejected. The acceptance threshold is higher than the grantmaker's preferred threshold. That is, the grantmaker commits to rejecting some applicants who the grantmaker would actually like to fund. This commitment to being tough generates powerful investment incentives.

The condition on the distribution of initial merit entails that the grantmaker's preferred threshold lie in the upper tail of the distribution.
In this sense, the grantmaker only wants to approve exceptional applicants and competition is fierce.

The optimal mechanism has an interesting alternative interpretation \citep[Section~5.3]{augias2023nonmarket}: instead of committing to being tough, the grantmaker allocates optimally based on a \emph{noisy} evaluation of final merit. One way of implementing the optimal mechanism via a noisy evaluation is as follows: if an applicant's final merit fails to pass the acceptance threshold, the evaluation only reveals that the merit failed to pass; else if the final merit passes the threshold, the evaluation perfectly reveals the final merit. The coarse evaluation of failure is important for circumventing the grantmaker's lack of commitment to being tough: since the acceptance threshold is strictly higher than the grantmaker's preferred threshold, a non-committed grantmaker would be tempted to approve applicants near the acceptance threshold, upsetting applicants' investment incentives. 

The alternative interpretation gives us a sense of how the grantmaker benefits from delegating the evaluation to referees who are tougher than the grantmaker.
The interpretation also speaks to the optimal level of noise: to stimulate productive investments, evaluation should be exact at the top but coarse at the bottom.

Let us briefly elaborate on why it is quite subtle that there is an optimal deterministic mechanism. Understanding the induced investment incentives is key. Given a deterministic threshold mechanism, applicants with initial merit narrowly below the threshold will invest to win a grant. 
However, applicants with initial merit above the threshold do not invest (they are sure to be funded without effort), and neither do applicants with initial merit far below the threshold (reaching the threshold is too costly).
The grantmaker could try to generate better investment incentives by approving \emph{randomly} for final merit levels near the threshold. If an applicant's initial merit is above the threshold, they are now incentivized to invest so as to escape the lottery. Of the applicants with low initial merit who previously did not invest, some are now incentivized to invest a little to become eligible for the lottery. The downside of approving randomly, of course, is that some applicants that the grantmaker would like to approve will be rejected. Further, some applicants who previously invested to push themselves above the threshold are discouraged from investing as they are now only rewarded randomly. 
The optimality of deterministic threshold mechanisms thus involves carefully analyzing how these forces balance.

\paragraph{Further reading}
There is a vast literature on contests studying how competitive forces shape incentives to exert effort.
This literature has a lot to say about how prize architecture---are there many small or few large prizes?---shapes investment incentives. The results are nuanced; see, for example, classic work by \citet{moldovanu2001optimal} or more recent work by \citet{fang2020turning}.
\citet{morgan2022limits} study the impact of evaluation noise on investment incentives, while \citet{kim2023choosing} study a setting where contestants choose their incentives and how noisily they are evaluated.
For further reading, see \citet{ContestsTheoryandTopics,vojnovic2015contest}.
An important caveat is that the contest literature typically focuses on the aggregate effort across all applicants, including those that go unfunded. This focus potentially limits the applicability to grant allocation since grantmakers may care more about the effort of those applicants who are funded than those who are unfunded, especially if the projects of unfunded researchers are never realized.

\subsection{Falsification}\label{sec:design_aspects:falsification}
Applicants may not only undertake productive investments in their proposals. They may instead spend considerable falsification efforts to make their proposals appear stronger than they are. 
This nonproductive effort is socially wasteful, hampers the grantmaker's ability to identify good applicants, and hurts the political sustainability of the grant-awarding institution. 

We discuss mechanisms that anticipate such falsification incentives, based on work by \citet{perez2022test}.
Roughly speaking,\footnote{We use Proposition 1 of \citet{perez2022test} to reinterpret their baseline model.} the model of \citet{perez2022test} coincides with that of \citet{augias2023nonmarket} except that the applicant's effort now only has an impact on the type observed by the grantmaker; the effort has no impact on the grantmaker's actual payoffs.

A high-level insight of \citet{perez2022test} is that optimal mechanisms involve \emph{productive falsification}. Even if everyone can craft a brilliant proposal, doing so will be easier for those with better research ideas. Applicants will give in to the temptation of falsifying their proposals, equipping the grantmaker with an (imperfect) instrument for inferring their underlying merit. 

Another insight is that, depending on the magnitude of the falsification costs, the optimal mechanism again involves a lottery. A lottery deters excessive falsification by lowering the reward from falsifying to a high level. An added welfare benefit is that fewer resources are devoted to wasteful falsification, which seems in line with common arguments in favor of lotteries (for example, \cite{gross2019contest}).

\paragraph{Further reading} Even though the grantmaker may benefit from productive falsification, \citet{perez2023fraud} note that mechanisms prone to falsification may be politically unsustainable and generate unfair advantages for applicants with higher gaming abilities. \citet{perez2023fraud} study optimal mechanisms that are immune to falsification attempts. \citet{li2024screening} study a related setup with multiple applicants and multiple grants. \citet{gross2019contest} study a related setup where the grantmaker lacks commitment power and argue that lotteries may improve welfare. \citet{myers2024tradeoffs} considers, among other things, a model based on \citet{gross2019contest} where falsification also has positive externalities.

\section{Post-award management}\label{sec:postaward_management}

While the prospective evaluation of applicants surely plays an important role, grantmakers and funded researchers continue interacting after the grant is conferred. Post-award management and retrospective evaluation expand the toolbox that grantmakers have at their disposal. \citet{goldstein2020know} document the extent of post-award management of grants at ARPA-E; see also \citet{azoulay2019funding}.

As suggested earlier, a concern in the post-award stage may be that grantmakers cannot monitor how researchers exactly use their funds (the hidden action problem).
We next illustrate, using the simple model of \citet{maurer2004procuring}, how retrospective evaluation helps mitigate this hidden action problem.\footnote{Maurer and Scotchmer discuss various instruments. We focus on their discussion of grants.}

\subsection{A simple model of post-award management}\label{sec:maurer_scotchmer}

The model of \citet{maurer2004procuring} unfolds over multiple periods. A researcher in each period needs a grant to carry out their project. When awarded a grant, the researcher obtains an immediate private benefit corresponding to, say, career advances or reputation. Working on the project is costly for the researcher but benefits society through knowledge production. The researcher can also choose to shirk by diverting the funds to other activities. No knowledge is produced whenever the research does not work on the project. 

Here, we have a classic hidden action problem: the grantmaker wants the researcher to work, but the researcher has an incentive to shirk as the costs are borne privately. 

The model also features a hidden information component. Researchers differ in how likely they are to develop fund-worthy research ideas in future periods. This productivity is privately known to the researcher but not to the grantmaker.

The grantmaker can retrospectively evaluate the researcher.
Specifically, at the end of each period, the grantmaker learns whether the researcher worked on the project, and can then decide whether to slash future funding.

This policy of retrospective evaluation is effective whenever the prospect of receiving future funding is sufficiently valuable to researchers. However, this is not the case for all researchers. Only those researchers who are sufficiently optimistic about having fund-worthy ideas in the future are willing to incur the costs of working today. For the other researchers, the threat of losing access to funds has little bite since they do not expect to have many fund-worthy ideas in the future.

The optimal grant-allocation policy in \citet{maurer2004procuring} has the following structure. The grantmaker assigns a grant to every first-time applicant. In the following periods, the grantmaker evaluates each researcher's past performance and only awards a grant to researchers who delivered results in the past. It follows that researchers with high productivity in generating ideas will work whenever they hold a grant and receive a grant whenever they apply for one. The other researchers will receive a grant only once and not deliver results. Thus, retrospective evaluation and repeated grant allocation jointly allow the grantmaker to channel funds to productive researchers and provide them incentives to use the grant effectively. In \Cref{box:retrospective_repeated}, we formalize these ideas.

\begin{modelbox}[label=box:retrospective_repeated]{Retrospective evaluation and repeated funding}
    In the following, we set up the model in \citet{maurer2004procuring} and show how retrospective evaluation paired with repeated funding can help align researchers' with grantmakers' incentives. 
    Researchers privately know the probability $\lambda$ of obtaining a fund-worthy idea in each period. 
    If a researcher works on their idea, they incur an effort cost $\kappa$. 
    To work on ideas, researchers need grants.
    A researcher's immediate private value for being funded is $v$.
    Researchers discount future payoffs at rate $r$.
    Under the rule where researchers are funded if and only if they worked in all past periods, a researcher of type $\lambda$ has an incentive to work in the first period if 
    \begin{align}\label{eq:MS:noshriking}
        \tag{No-Shirking}
       v \leq v - \kappa + \frac{\lambda}{r}(v - \kappa).
    \end{align}
    By shirking (on the left side), the researcher enjoys the private benefit $v$ at zero cost but loses out on all future net benefits. By working, they pay the cost $\kappa$; additionally, they enjoy future benefits from funding equal to the per-idea rent $v - \kappa$ times the probability $\lambda$ of obtaining an idea in any given period, in perpetuity from the next period (and, thus, divided by the discount rate $r$). The \eqref{eq:MS:noshriking}-condition can be rearranged to
    \begin{align*}
       \lambda \geq \frac{r\kappa}{v - \kappa}.
    \end{align*}
    This inequality shows that only sufficiently productive researchers (high $\lambda$) have the incentives to work hard for continuous funding.
\end{modelbox}

\subsection{The pre-award benefit of post-award management}

We just saw that post-award management impacts the earlier allocation stage.
We now elaborate on this idea using work by \citet{mylovanov2017optimal}.

A grantmaker allocates one grant among a group of applicants. Each applicant privately knows their own individual merit. 
After allocating the grant, the grantmaker learns the funded applicant's merit and can impose a limited penalty. 
Our interpretation is that the grantmaker can infer the applicant's true merit by observing, say, the number of their publications. To penalize the applicant, the grantmaker can debar them from future calls or terminate the grant prematurely. However, penalization is imperfect.
The winner may enjoy some career gain from winning the grant, which the grantmaker cannot undo ex-post.
Premature termination has bite when researchers rely on the grant to carry out their work or when they are intrinsically motivated about working on their projects. This feature contrasts with the Maurer-Scotchmer model; there, researchers only care about getting funding.

How can the grantmaker use the threat of ex-post penalization to screen the applicants?
\citet[Theorem 1]{mylovanov2017optimal} derive the following optimal mechanism, taking the form of a \emph{binary shortlisting procedure}.\footnote{Here, we focus on the optimal mechanism in the case with sufficiently many agents. See Theorem 1 in \citet{mylovanov2017optimal} for the general solution.}
First, the grantmaker announces a threshold. Second, applicants are asked to declare whether their merit lies above the threshold (and they will find it in their best interest to do so truthfully). Third, applicants above the threshold are shortlisted, and applicants below the threshold are shortlisted with some probability strictly less than one. 
Fourth, a winner is selected uniformly at random from the shortlist (if the shortlist is empty, a winner is selected uniformly at random from the full applicant pool). Finally, if the winner is ex-post found to have misreported their quality, the penalty is triggered.

We highlight two qualitative insights from this result.
First, the optimal mechanism involves a lottery. To gain an intuition, consider applicants with merit below the threshold. 
Applicants are more likely to be shortlisted (and hence win the grant) if they claim to pass the threshold.
Therefore, if a low-merit applicant would never win the grant by reporting truthfully, they would have an incentive to misreport, even under the threat of ex-post penalization. 
By introducing the lottery, the grantmaker incentivizes these low-merit types to report truthfully. Thus, the lottery serves as an incentive device for ensuring truthful reports.

The second insight is that ex-post penalties indeed provide incentives ex-ante.
Applicants are more likely to be short-listed (and hence win the grant) if they pass the threshold.
Thus the optimal mechanism is more likely to allocate the grant to a high-merit researcher than a completely random mechanism.

\paragraph{Further reading} \citet{li2020mechanism} studies a nearby setting where the grantmaker has to pay a cost to learn the winner's type ex-post.

\subsection{Post-award incentives}\label{sec:design_of_grants:postallocation}
We next ask how the structure of grants shapes the incentives of funded researchers. As we have seen in \Cref{sec:maurer_scotchmer}, the nature of grants as upfront payments creates a hidden action problem: conditional on being funded, a researcher may have an incentive to use the funds to serve their own interests rather than those of the grantmaker or society. \citet{maurer2004procuring} emphasize that grantmakers can re-align interests by monitoring researchers and, possibly, cutting them off from future funding. However, slashing funding might not be a sufficiently powerful instrument. Researchers can apply for grants from different grantmakers or might not require external funding frequently. Further, it is somewhat stringent to assume that grantmakers can perfectly monitor whether researchers work.

In classical hidden action models \citep{holmstrom1979moral}, incentives are typically provided through performance-dependent payments. Grants do not usually use outcome-dependent rewards, but we can still gain some insights from this literature. In the following, we will first give an overview of relevant insights from the literature assuming that performance-dependent rewards are available to the funder. Then, we will briefly discuss how such performance-dependent rewards can be interpreted in the context of research grants.

Consider a researcher who has been awarded a grant, and there is no uncertainty about the researcher's merit. Once the grant has been allocated, the researcher must decide how to allocate their resources; for example, whether to spend time and effort on the projects in their proposal. While the grantmaker wants the researcher to devote time to those projects, the researcher has private incentives not to; for example, because the researcher prefers working on alternative projects or traveling to conferences. Generally, a grantmaker cannot observe or verify how a researcher allocates their time. Therefore, the grantmaker must use verifiable outputs that depend on the researcher's choices to provide incentives, such as publications related to the grant projects. Due to the uncertain nature of research, publication outcomes are not a perfect measure of the researcher's choices, but they are likely correlated. Suppose that, by allocating more time to grant-related projects, a researcher can increase the likelihood of having publications at the end of the grant period. Then, the grantmaker can design a reward scheme for the researcher that depends on publication outcomes. Good publication outcomes serve as a signal of the researcher's having allocated their time as the grantmaker desired and are therefore rewarded. If a researcher has worse publication outcomes, the reward should be lower to prevent the researcher from allocating their time differently.

This basic hidden action model has been extended in many directions. In the context of scientific research, we focus on two relevant modeling ingredients. First, grants usually have long time horizons, and researchers repeatedly decide how to allocate their resources. Second, researchers learn about their projects over time---for example, the feasibility of producing any (publishable) results. 

For simplicity, suppose the grantmaker wants to encourage a researcher to work on a project for a specified duration. The successful completion is observable (in the form of journal publications, for example). However, whether and when the researcher successfully completes the project is uncertain, even when working on the project as desired. Therefore, the grantmaker cannot infer whether the absence of a publication at each point is due to misfortune or idleness. How can the grantmaker incentivize the researcher to work on the project?

Let us consider the model of \citet{green2016breakthroughs}. The grantmaker desires that the researcher successfully completes the funded project and chooses both the duration of the project and how the researcher is rewarded. The researcher then chooses at each point in time whether or not to work on the project. When working on the project, the researcher forgoes the value of using the funds differently. Naturally, the project can only be completed when the researcher works on the project. To capture the uncertainty in research, the researcher may fail to complete the project even when working on it. \citet{green2016breakthroughs} show that the optimal grant design in this case is as follows. The grant provides funding until a (finite) deadline. The reward for completing the project is time-dependent; the earlier the researcher completes the project, the higher the reward. Intuitively, the researcher has to be compensated for two reasons. First, when working on the project, the researcher forgoes the benefits of allocating the funds differently. Second, upon completing the project before the deadline, the researcher forgoes the private enjoyment from the funds for the remaining time---this is because the researcher could always decide to use the funds differently and claim that the absence of success was due to bad luck. 

To capture the idea that a researcher learns about a project while working on it, consider the following model studied by \citet{moroni2022experimentation}. It is uncertain whether the project is feasible. If the project is infeasible, it cannot be completed. Thus, the researcher learns about the project's feasibility while working. Specifically, when working without success, the researcher becomes more pessimistic about the project, and hence also more pessimistic about the prospect of winning whatever reward the grantmaker promised for successful completion. The more pessimistic the researcher is, the higher the reward necessary to incentivize effort. Therefore, one might expect the reward to increase over time in the presence of learning. However, this is not the case. Suppose that the researcher would expect a higher reward tomorrow than today. Then, the researcher would rather delay working on the project today (which is unobserved by the grantmaker) and work on the project tomorrow. Because the researcher did not work today, the researcher did not become more pessimistic about the project. Therefore, the researcher must be compensated for earlier successes. The optimal design features a project deadline and a constant reward.\footnote{The constant reward relies on the agent's not discounting the future. If the agent would discount the future, the optimal reward schedule would be declining over time as the incentive to delay the work is slightly mitigated.} 

\citet{halac2016optimal} show how to incorporate hidden information into this model. Suppose the researcher has better information about the project's feasibility than the grantmaker. In this case, the grantmaker can offer a menu of different grant designs so that the researcher reveals their private information via self-selection. \citet{green2016breakthroughs} show how the optimal grant design changes if completing the project requires an intermediate step that is unverifiable to the grantmaker. In this case, the grantmaker optimally incentivizes the researcher by staging the project into two steps with a soft first and a hard second deadline. 

\citet{manso2011motivating} considers a different variant of the dynamic hidden action model focusing on the researcher's choice between two projects. One project is safe while the other project is riskier; in particular, it is unknown how likely it is to succeed on the risky project, and, in expectation, success is less likely than on the safe project. \citet{manso2011motivating} shows that to encourage work on the riskier project, the grant design must exhibit tolerance for early failure (reflected, for example, in a longer duration) and rewards for long-term success. If early successes are rewarded relatively highly, the researcher is incentivized to work on safe, less creative problems rather than novel but risky problems.

On the empirical side, \citet{myers2023money}, using survey data of US professors, provide evidence that a longer grant duration indeed encourages riskier projects, but only for researchers with long-term job security.\footnote{Here, when we say a grant has a size of $\$ x$ and a duration of $n$ years, we mean that the grantee can freely spend up to $\$ x$ within $n$ years. After $n$ years, all unused funds expire.} In response to larger grants researchers shift resources toward existing projects rather than opening new ones. \citet{myers2023money} also discuss estimates for the rate at which researchers are willing to substitute duration for size.\footnote{See \citet{tham2023science,tham2024scientific} for the impact of funding delays on career outcomes.}

The results in the preceding discussion of hidden action models rely on the ability of the grantmaker to design and commit to performance-dependent rewards. Whenever such rewards are infeasible or undesirable, the results provide a benchmark on what would be achievable with these additional instruments. It would be interesting to investigate how a grantmaker can design instruments to resemble performance-dependent payments. \citet{maurer2004procuring} provide one example of repeatedly assigning grants of fixed value. One could imagine that the grantmaker can condition the access to or the design of future grants on past performance to approximate the optimal performance-dependent rewards. \citet{gross2023rationalizing} interpret the reward in a moral hazard model normatively as how the scientific community \emph{should} evaluate scientific contributions.

Another strand of the literature on dynamic hidden action models focuses on the use of costly inspections to incentivize researchers to work. \citet{ball2023should} consider a dynamic model in which a researcher decides whether to work on a project. The grantmaker cannot use time-dependent payments in their model but chooses the timing of costly inspections that reveal the researcher's past choices. The researcher's effort increases the probability that a project is completed by working. Completing the project terminates the grant. How can the grantmaker time inspections in such a way that the researcher will work while economizing on the cost of inspections? \citet{ball2023should} show that the grantmaker optimally inspects the researcher in fixed time intervals.\footnote{\citet{ball2023should} also study the case in which working reduces the probability of a \emph{breakdown}, that is, the failure of the project. In this case, random inspections are optimal.}

\section{Grant supply and the direction of research}\label{sec:supply_and_direction}
The previous sections shed light on \emph{how} the grantmaker should structure payments and design the allocation rule to achieve a particular goal.
We now shift our focus to the choice of \emph{what} to fund. How does a grantmaker's funding of a specific topic impact researchers' incentives and the direction of research? 

In practice, grantmakers frequently offer grants on specific topics, so-called mission-oriented grants.
Examples include \emph{Requests for Applications (RFAs)} at the NIH \citep{myers2020elasticity}, and the \emph{SWITCHES} program at ARPA-E \citep{azoulay2020scientificgrantfunding}. 
Further, in the context of investigator-initiated proposals, grantmakers can selectively fund proposals that align most closely with their goals, thereby indirectly steering the direction of research. Investigator-initiated proposals perhaps run a higher risk than mission-oriented grants of attracting proposals that do not closely fit the grantmaker's interests. However, investigator-initiated proposals can more easily leverage researchers' knowledge about the prospects of research projects.

Besides the availability of funding and personal intellectual curiosity, researchers' preferences over questions are shaped by feasibility---how difficult is progress given current knowledge?--- and the reward for establishing priority. Importantly, the incentives from feasibility and priority may fail to induce an efficient allocation of researchers to questions due to \emph{congestion} and \emph{competition} effects.

When a new researcher joins a field, the rate of discovery in this field increases. \emph{Congestion effects} arise when this increase is decreasing in the number of incumbent researchers. From a social planner's perspective, it may be preferable for the new researcher to reallocate to a field with fewer incumbent researchers. In theoretical work, \citet{hopenhayn2021direction} develop this idea in a model of corporate R\&D.
Since firms cannot acquire property rights to open problems (but only to inventions), there is inefficiently high entry into ``hot'' areas. While theirs is a model of patents and corporate R\&D, a similar line of reasoning may apply to academic research since researchers cannot acquire property rights to open questions. Thus, there may be an economic value to communal norms that informally bestow such property rights. 

For projects with high ex-ante potential, \emph{competition effects} arise when many researcher simultaneously tackle the problem and attempt to outrace one another to establish priority.
\citet{hill2024race} model these incentives and provide evidence of a negative effect on research quality, finding that projects with ``higher ex-ante potential generate more competition, are completed faster, and are lower quality.''\footnote{Their model does not assume that establishing priority is a winner-takes-all-contest, but only that being second yields a lower reward than being first; this aligns with empirical evidence from structural biology \citep{hill2023scooped}. The model of \citet{hill2024race} builds on earlier theoretical work by \citet{bobtcheff2017researcher}, who cast competition between two researchers as a winner-takes-all contest without endogenous entry. \citet{bobtcheff2017researcher} show that ``when breakthroughs become more frequent, researchers are under increasing competitive pressure and have decreasing incentives to wait and let their ideas mature.'' That is, the adverse incentives to publish low-quality work to establish priority are especially pertinent when technological progress allows the research community to produce insights quickly.}

How can grants improve the allocation of researchers to research questions and curtail congestion and competition effects? Let us speculate on some channels. First, grantmakers can create ``hot'' areas by offering large grants, thereby pulling researchers away from congested areas. Naturally, implementing this would require an idea of what researchers would otherwise work, an issue we return to momentarily. Second, if unfunded researchers cannot investigate certain questions (due to the costs of specialized equipment, for example), grantmakers effectively place an upper bound on the number of researchers working on a question, reducing both competition and congestion effects. Third, grants may act as a commitment device: a researcher with a grant credibly signals that they have the resources to investigate their area.

How effective are grants at guiding the direction of research? Recent empirical work by \citet{myers2020elasticity} quantifies the costs of steering the direction using RFAs at the NIH. The evidence suggests that researchers do not simply follow the money, as it were, but are much more likely to apply for RFAs that are topically similar to their prior research. Overall, \citet{myers2020elasticity} suggests it is costly to incentivize specific topics. In equilibrium, ``RFAs must make more funds available to attract the same number of applications as the investigator-initiated mechanism.''

The costs and benefits of investigating a question are difficult to quantify since they depend on the current knowledge frontier. \citet{carnehl2021quest} develop a model for understanding the evolution of scientific knowledge and the researchers' incentives for pursuing questions.
The model conceptualizes knowledge as a set of questions to which the answer is known. The answer to any particular question has a spillover effect on nearby questions through improved conjectures about their answers. Having precise information about answers is valuable since it guides decision-making in practical problems. 
In this model, the interests of researchers and society are misaligned for two reasons. First, researchers do not fully internalize that uncovering an answer provides a guiding light for future researchers. Second, researchers bear the costs of uncovering answers; these costs are lower for questions similar to ones with known answers. The misaligned interests lead to inefficiently low novelty in research. Hence this model provides a formal framework for investigating the effects of policy tools (such as RFAs) when researchers have a stronger preference than society for questions close to existing knowledge. 
Interpreting grants as reductions in the cost of research, grants can induce researchers to investigate more novel questions, even when grants are not topic-specific. In this model, however, such grants cannot incentivize ``moonshots''---research on extremely novel questions that guide future researchers to choose more efficiently. To encourage researchers to work on moonshot questions, one may have to resort to other tools, such as research prizes or mission-oriented grants.

%% file: concluding_remarks.tex
\section{Concluding remarks}\label{sec:concluding_remarks}
Two high-level insights that emerge from our overview of the economics of grant design:
\begin{enumerate}
    \item The case for (partial) randomization is mixed. For example, randomization may arise optimally if grantmakers are concerned with evaluation costs or with deterring applicants from inflating their merit. The case for randomization is weaker when accounting for productive investments.
    \item Intermediate reviews and a staged grant design may help realign the incentives of funded researchers with those of the grantmaker, ensuring that researchers direct their resources toward productive areas.
\end{enumerate}

The economics literature has given relatively little attention to grant funding. By collecting insights that are broadly applicable to allocation problems, we hope this piece inspires follow-up work on grant funding as a particular application. In our view, there is scope for both theoretical and empirical work clarifying the interaction of different stages and economic forces in grant funding (\Cref{fig:intro_grant_funding,fig:grantfunding}), as well as improving our understanding of the relationship between grants and other funding instruments (\Cref{fig:intro_science_funding}). Let us give two examples.
\begin{enumerate}
    \item We discussed separately the role of applicants' productive and unproductive efforts.
    One could think that efforts are rarely completely unproductive.
    Indeed, \citet{myers2024tradeoffs} uses survey data to argue that applicants' fundraising efforts typically do have some positive externalities.
    It is interesting to analyze in more detail the application and evaluation costs inherent to grant funding, and to then draw lessons for the design of funding rules.
    \item As already indicated, there is little prior work relating grants to other funding instruments or, more broadly, relating push to pull incentives in science funding. 
    The COVID-19 pandemic spurred greater interest in the question of optimally combining push and pull incentives, perhaps via entirely novel instruments \citep{souza2024ows,athey2022expanding,ahuja2021preparing}.
\end{enumerate}